\documentclass{elsart3p}

\usepackage[square, comma, sort & compress]{natbib}

 \usepackage{graphicx}
\usepackage{lineno}
\usepackage{subfigure}
\usepackage{multirow}

\usepackage{amssymb}
\journal{Nuclear Instruments and Methods A}
\begin{document}

\begin{frontmatter}

\vspace{-0.85 in}

\title{An array of low-background $^3$He proportional counters \\ for the Sudbury Neutrino Observatory}
\author[UW]{J. F. Amsbaugh},
\author[LANL]{J. M. Anaya},
\author[LANL]{J. Banar},
\author[LANL]{T. J. Bowles},
\author[UW]{M. C. Browne},
\author[UW]{T. V. Bullard},
\author[UW]{T. H. Burritt},
\author[UW]{G. A. Cox-Mobrand},
\author[Oxford]{X. Dai},
\author[Penn]{H. Deng},
\author[Queen]{M. Di Marco},
\author[UW]{P. J. Doe},
\author[LANL]{M. R. Dragowsky},
\author[UW]{C. A. Duba},
\author[Queen]{F. A. Duncan},
\author[Queen]{E. D. Earle},
\author[LANL]{S. R. Elliott $^{\rm a, \hspace{-0.75 mm}}$},
\author[LANL]{E.-I. Esch},
\author[Oxford]{H. Fergani},
\author[UW]{J. A. Formaggio},
\author[LANL]{M. M. Fowler},
\author[UW]{J. E. Franklin},
\author[UW]{P. Geissb\"{u}hler},
\author[LANL]{J. V. Germani $^{\rm a, \hspace{-0.75 mm}}$},
\author[LANL]{A. Goldschmidt},
\author[Queen]{E. Guillian},
\author[Queen]{A. L. Hallin},
\author[UW]{G. Harper},
\author[Queen]{P. J. Harvey},
\author[UW]{R. Hazama},
\author[UW]{K. M. Heeger},
\author[Queen]{J. Heise $^{\rm b, \hspace{-0.75 mm}}$},
\author[LANL]{A. Hime},
\author[UW]{M. A. Howe},
\author[Texas]{M. Huang},
\author[Queen]{L. L. Kormos},
\author[Queen]{C. Kraus},
\author[Queen]{C. B. Krauss},
\author[Guelph]{J. Law},
\author[SNO]{I. T. Lawson $^{\rm g, \hspace{-0.75 mm}}$},
\author[UC]{K. T. Lesko},
\author[UC]{J. C. Loach $^{\rm c, \hspace{-0.75 mm}}$},
\author[Oxford]{S. Majerus},
\author[UW]{J. Manor},
\author[UW]{S. McGee},
\author[UW]{K. K. S. Miknaitis},
\author[LANL]{G. G. Miller},
\author[SNO]{B. Morissette},
\author[UW]{A. Myers},
\author[UW]{N. S. Oblath},
\author[Oxford]{H. M. O'Keeffe},
\author[LANL]{R. W. Ollerhead $^{\rm g, \hspace{-0.75 mm}}$},
\author[Oxford]{S. J. M. Peeters},
\author[UC]{A. W. P. Poon},
\author[UC]{G. Prior},
\author[Guelph]{S. D. Reitzner},
\author[LANL]{K. Rielage $^{\rm a, \hspace{-0.75 mm}}$},
\author[UW]{R. G. H. Robertson},
\author[Queen]{P. Skensved},
\author[UC]{A. R. Smith},
\author[LANL]{M. W. E. Smith $^{\rm a, \hspace{-0.75 mm}}$},
\author[UW]{T. D. Steiger},
\author[LANL]{L. C. Stonehill $^{\rm a, \hspace{-2 mm}}$ \corauthref{cor}},  
\corauth[cor]{Corresponding author. Telephone: 505-665-3821.  E-mail address: lauracs@lanl.gov.}
\author[UW]{P. M. Thornewell $^{\rm b, c, \hspace{-0.75 mm}}$},
\author[UC]{N. Tolich},
\author[UW]{B. A. VanDevender},
\author[UW]{T. D. Van Wechel},
\author[UW]{B. L. Wall},
\author[UW]{H. Wan Chan Tseung $^{\rm c, \hspace{-0.75 mm}}$},
\author[UBC]{J. Wendland},
\author[Oxford]{N. West},
\author[LANL]{J. B. Wilhelmy},
\author[UW]{J. F. Wilkerson},
\author[LANL]{J. M. Wouters}
\address[UW]{Center for Experimental Nuclear Physics and Astrophysics, and Department of Physics, University of Washington, Seattle, WA 98195, USA}
\address[LANL]{Los Alamos National Laboratory, Los Alamos, NM 87545, USA}
\address[Oxford]{Department of Physics, University of Oxford, Oxford OX1 3RH, UK}
\address[Penn]{Department of Physics and Astronomy, University of Pennsylvania, Philadelphia, PA 19104, USA}
\address[Queen]{Department of Physics, Queen's University, Kingston, Ontario K7L 3N6, Canada}
\address[Texas]{Department of Physics, University of Texas at Austin, Austin, TX 78712, USA}
\address[Guelph]{Physics Department, University of Guelph, Guelph, Ontario N1G 2W1, Canada}
\address[SNO]{SNOLAB, Lively, Ontario P3Y 1M3, Canada}
\address[UC]{Institute for Nuclear and Particle Astrophysics, and Nuclear Science Division, Lawrence Berkeley National Laboratory, Berkeley, CA 94720, USA}
\address[UBC]{Department of Physics and Astronomy, University of British Columbia, Vancouver, BC V6T 1Z1, Canada}

\begin{abstract}

An array of Neutral-Current Detectors (NCDs) has been built in order to make a unique measurement of the total active flux of solar neutrinos in the Sudbury Neutrino Observatory (SNO).  Data in the third phase of the SNO experiment were collected between November 2004 and November 2006, after the NCD array was added to improve the neutral-current sensitivity of the SNO detector.  This array consisted of 36 strings of proportional counters filled with a mixture of $^3$He and CF$_4$ gas capable of detecting the neutrons liberated by the neutrino-deuteron neutral current reaction in the D$_2$O, and four strings filled with a mixture of $^4$He and CF$_4$ gas for background measurements.  The proportional counter diameter is 5 cm.  The total deployed array length was 398 m.  The SNO NCD array is the lowest-radioactivity large array of proportional counters ever produced.  This article describes the design, construction, deployment, and characterization of the NCD array, discusses the electronics and data acquisition system, and considers event signatures and backgrounds.  

\end{abstract}

\end{frontmatter}


\section{The Sudbury Neutrino Observatory}

The Sudbury Neutrino Observatory (SNO) is the first solar neutrino detector capable of measuring the flux of electron-type solar neutrinos above 5~MeV as well as the total flux of all active solar neutrinos above 2.2 MeV.  In order to reduce cosmic-ray backgrounds, SNO is located 2092 m underground in the Creighton Nickel Mine owned by CVRD Inco, Ltd. near Sudbury, Ontario, Canada.  The central volume of SNO is 1000 metric tonnes of ultrapure heavy water (D$_2$O) contained in an acrylic vessel (AV).  Strict controls of radioactive impurities were applied to all materials used in the SNO detector, to minimize backgrounds and allow extraction of the neutrino signal of approximately ten events  per day.  Neutrinos are detected in SNO through the reactions:

\begin{equation}
\rm \nu_x + e^-  \rightarrow \nu_x + e^-  \\
\end{equation}
\begin{equation}
\rm \nu_e + d \rightarrow e^- + p + p - 1.44~ MeV  \\
\end{equation}
\begin{equation}
\rm \nu_x + d \rightarrow \nu_x + n + p - 2.22~ MeV
\end{equation}
where $\rm \nu_x$ refers to any active flavor of neutrino.  SNO can detect the first two interactions when an electron produces Cherenkov light that reaches the photomultiplier tube (PMT) array surrounding the D$_2$O.  The elastic scattering (ES) of electrons by neutrinos (Eq. 1) is sensitive mostly to electron-type neutrinos.  These events are highly directional, establishing the solar origin of the neutrinos.  The charged-current (CC) absorption of an electron-type neutrino by the deuteron (Eq. 2) produces an electron with an energy related to that of the incident neutrino, allowing SNO to measure the energy spectrum of these neutrinos above an analysis threshold of $\sim 5$~MeV.  The neutral-current (NC) disintegration of a deuteron by a neutrino of any active flavor produces a neutron and a proton (Eq. 3) and has an energy threshold of 2.2 MeV.  SNO was designed to detect the neutron liberated from this reaction in three different ways, each in its own phase of the experiment.  A complete description of the SNO detector has been published previously \cite{SNOnim}.

In the first phase of SNO, the D$_2$O phase, the thermalized neutron was absorbed by a deuteron with a 0.5-mb cross-section \cite{crossSec}, releasing a 6.25-MeV photon.  The photon Compton scattered, imparting enough energy to the electrons to create Cherenkov light that was detected in the PMT array.  The Cherenkov light from these neutron captures was statistically separated from the CC and ES Cherenkov signals using distributions of the reconstructed energy, position, and direction of the events.  This phase of SNO ran from November 2, 1999 to May 31, 2001, and results have been published \cite{d2o1,d2o2,d2o3,longd2o}.  

For the second phase, the salt phase, two tonnes of purified NaCl were added to the D$_2$O.  This additive enhanced the probability of neutron capture within the central volume because the 44-b thermal capture cross-section on $^{35}$Cl is more than 80,000 times larger than the cross-section on deuterium \cite{crossSec}.  Absorption of a neutron on $^{35}$Cl produces an 8.6-MeV cascade of photons that Compton scatter, yielding Cherenkov light that was detected with the PMT array.  Compared to the D$_2$O phase, the larger isotropy of the light from this cascade allowed for a more accurate separation of the NC events from those produced by other neutrino reactions.  A complete analysis of results from this phase has been performed \cite{salt1,salt2} and encompasses data from July 26, 2001 to August 28, 2003.

The third and final phase, the NCD phase, utilized an array of proportional counters filled with a $^3$He-CF$_4$ gas mixture for neutron detection independent of the PMTs.  These Neutral-Current Detectors (NCDs) capture the neutron released in Eq. 3, thereby measuring the rate of this reaction.  Though more technically complicated than the D$_2$O or salt methods, the use of NCDs has many advantages.  The NCDs were distributed in the D$_2$O volume, and only blocked 9\% of the Cherenkov light.  Over 60\% of the detected NC events were recorded separately from the CC and ES signals and can be distinguished on an event-by-event basis.  Thus the NC flux can be measured without performing the statistical separation that was necessary in the previous phases of SNO, reducing the correlation between the NC and CC measurements from about $-0.5$ to better than $-0.02$.  This phase is expected to produce a result for the NC flux of comparable precision to previous phases of SNO, but with very different systematics.  Additionally, the CC signal in the NCD phase has substantially reduced contamination from neutron capture.  The NCD phase provides an opportunity to confirm the previous results from SNO in an `independent' detector.

Installation of the NCD array in the SNO detector began in November 2003 after the NaCl was removed, and was completed in April 2004.  After a period of commissioning, data-taking with the complete detector system began on November 27, 2004 and concluded on November 28, 2006.  This article discusses the design and construction of the NCDs and their deployment into the SNO detector.  The NCD electronics and data acquisition system are also described, as well as the signals and backgrounds in the NCDs and results from radiopurity measurements of the NCD array.  

\section{The SNO Neutral-Current Detectors}
\label{SNONCD}

Standard Solar Models (SSM) \cite{ssm,ssm2, ssmTC, ssm3} predict that approximately 13 neutrons are produced per day by NC interactions within SNO's heavy water. It is essential that care be taken to minimize backgrounds if such small rates are to be reliably measured. Proportional counters are well-suited for extremely low-background applications since the gaseous active medium can be readily purified.  The background level of proportional counters is set by cosmogenics, radioactivity in construction materials, and spurious pulses originating from sources other than ionizing radiation.

Since D$_2$O is an excellent moderator with low absorption of thermal neutrons, it is possible to detect free neutrons by placing a sparse array of proportional counters filled with a neutron-absorbing gas in the heavy water.  It was decided to fill the proportional counters with $^3$He, since its use for neutron detection is well-established \cite{3He} and takes advantage of its large neutron-capture cross section of 5333 b \cite{crossSec}.  Neutron detection in a $^3$He proportional counter occurs via the reaction:
\begin{equation}
\rm ^3He + n \rightarrow p + ^{3}\!H + 764~ keV
\end{equation}
which produces a back-to-back proton-triton pair, with the proton carrying 573 keV of kinetic energy and the triton having 191 keV.  These energetic ions create over 20,000 electron-ion pairs.  The electrons then drift towards a central anode wire, where they undergo avalanche multiplication resulting in a large number of secondary electron-ion pairs that produce an electrical signal on the anode.  

\begin{figure}[htb]
\begin{center}
\includegraphics*[width=\columnwidth]{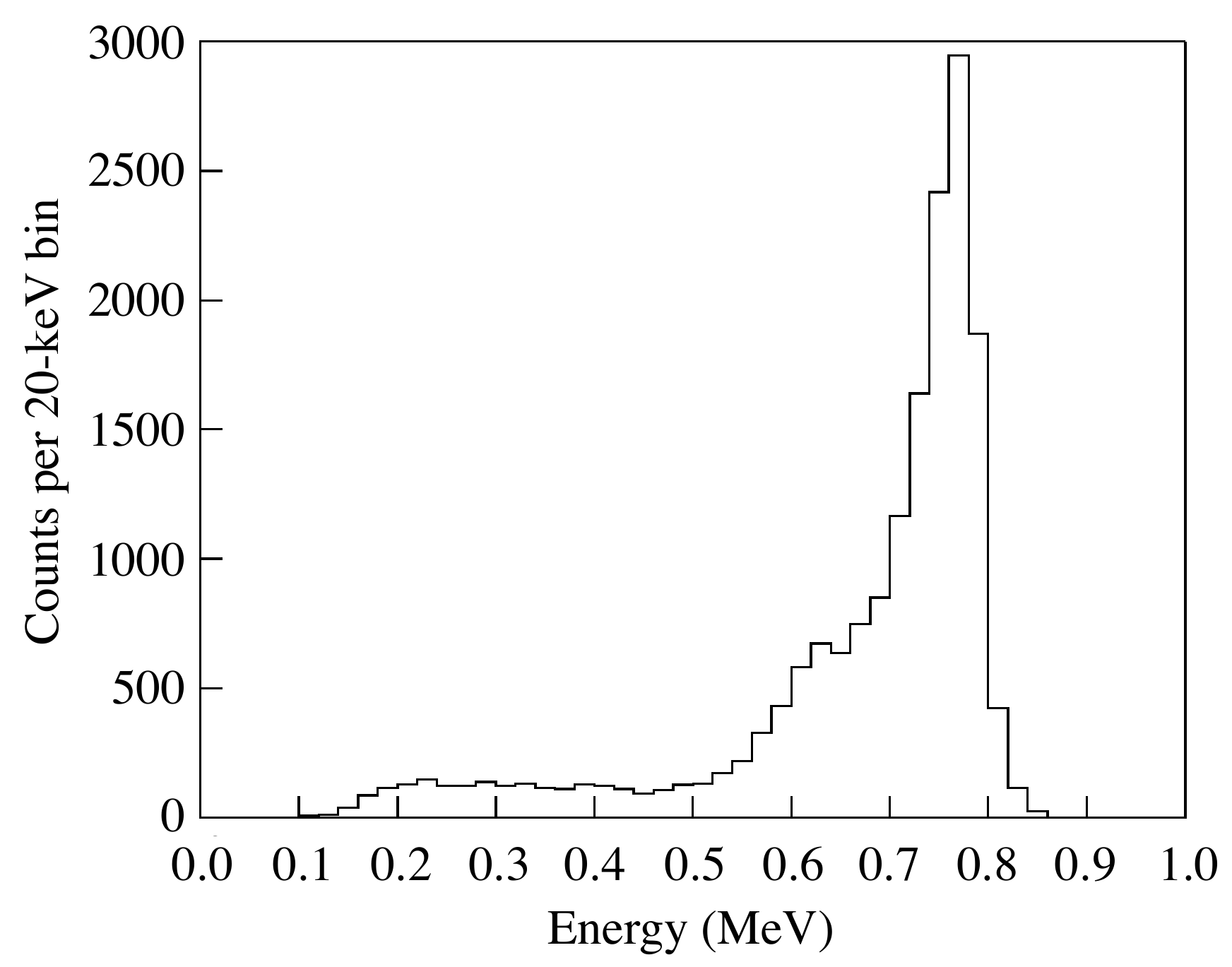}
\caption[wall effect]{NCD array neutron-capture spectrum from a uniformly-distributed $^{24}$Na calibration.  The peak at 764 keV corresponds to deposition of the full kinetic energy of the proton and triton in the active volume of the NCD.  The 573-keV shoulder, caused by total absorption of the triton's energy in the wall, is distorted by space-charge effects, discussed in Section~\ref{signals}.  The 191-keV shoulder is caused by total absorption of the proton's energy in the wall.}
\label{spect}
\end{center}
\bigskip
\end{figure}

Despite the discrete energy, the neutron spectrum of a $^3$He proportional counter is not simply a sharp peak at 764~keV.  When a neutron capture occurs near the wall of the proportional counter, the proton or triton may strike the wall before stopping in the gas, causing events that leave less than 764~keV in the gas.  An energy deposition of 191~keV corresponds to total absorption of the proton's energy in the wall \cite{wall}.  Energy depositions below 191 keV are extremely rare (less than 0.1\% of the events), due to the limited geometric phase space for both particles to strike the wall.  Figure \ref{spect} shows a neutron-capture energy spectrum from the NCD array, illustrating this `wall effect', which can be decreased by adding another component to the gas that contributes stopping power.  CF$_4$ was used in the NCDs for this purpose, and to provide quenching.

There are several potential sources of background events that may occur in the 191 to 764 keV energy region.  Nuclei that are embedded in the NCD walls and on the inner surface emit alpha particles that produce ionizing events with a broad spectrum of energies, including the neutron-capture energy region of interest.  Compton electrons, photoelectrons, and electrons from beta decays are a less serious problem because extensive multiple scattering is usually required before such processes can deposit enough energy in the gas to mimic a neutron event.  However, tritium contamination in the $^3$He must be kept to low levels (on the order of nCi/STP-L) by purification to avoid pile-up of betas from tritium decay.  If tritium is controlled, then the chief source of background to the neutron-capture signal in $^3$He proportional counters is the natural alpha emitters.

In the case of SNO, neutrons produced by any process other than the NC interaction of neutrinos with deuterium constitute an additional background.  Gamma rays above the 2.2-MeV breakup threshold of the deuteron can induce photodisintegration that liberates a free neutron.  Both thorium and uranium must be controlled because of the high-energy gammas emitted near the bottom of both decay chains (2.614 MeV from $^{208}$Tl and several gammas above 2.2 MeV from $^{214}$Bi, with the most probable being 2.448 MeV).   Because the branching fraction to the high-energy gamma is approximately 36\% in the $^{232}$Th chain, but only about 2\% in the $^{238}$U chain, more stringent limits must be placed on the amount of $^{232}$Th.  A contamination of 3.8 $\mu$g of $^{238}$U or 0.5 $\mu$g of $^{232}$Th in the NCD array corresponds to 1\% of the NC neutron production rate.  The combined limit applied in the design of the NCD array was 2\%.  Commercial proportional counters could not be used, since they did not meet these stringent radioactivity limits.  To reach the required cleanliness, the NCDs were designed and built by the SNO collaboration using low-radioactivity materials.

\begin{figure}[bp]
\begin{center}
\includegraphics*[width=0.87\columnwidth]{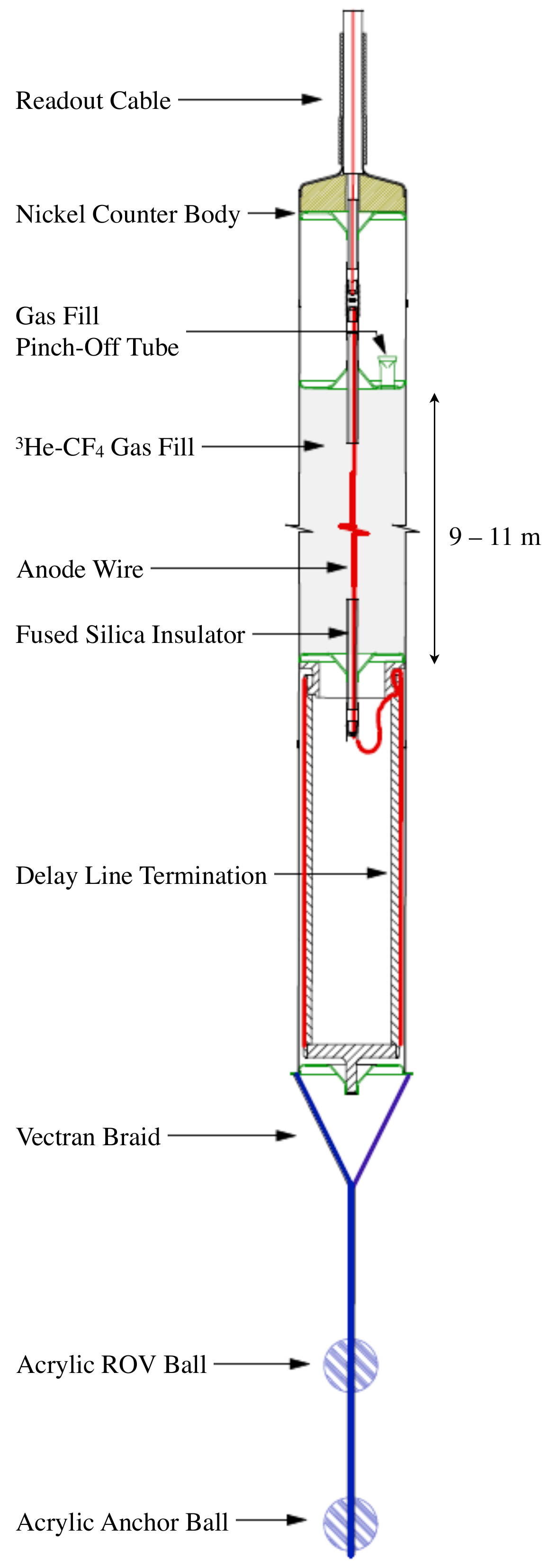}
\caption[NCD]{Schematic of an NCD string with readout cable, active region, delay line, and anchor system.  The 9 -- 11 meter long active region (not to scale) consisted of three or four individual gas volumes that were electrically connected and laser welded (these connections are shown in Figure \ref{fig:endcap} and described in Section \ref{endcapSection}).}
\label{ncd}
\end{center}
\bigskip
\end{figure}

\begin{figure}[htb]
\begin{center}
\includegraphics*[width=\columnwidth]{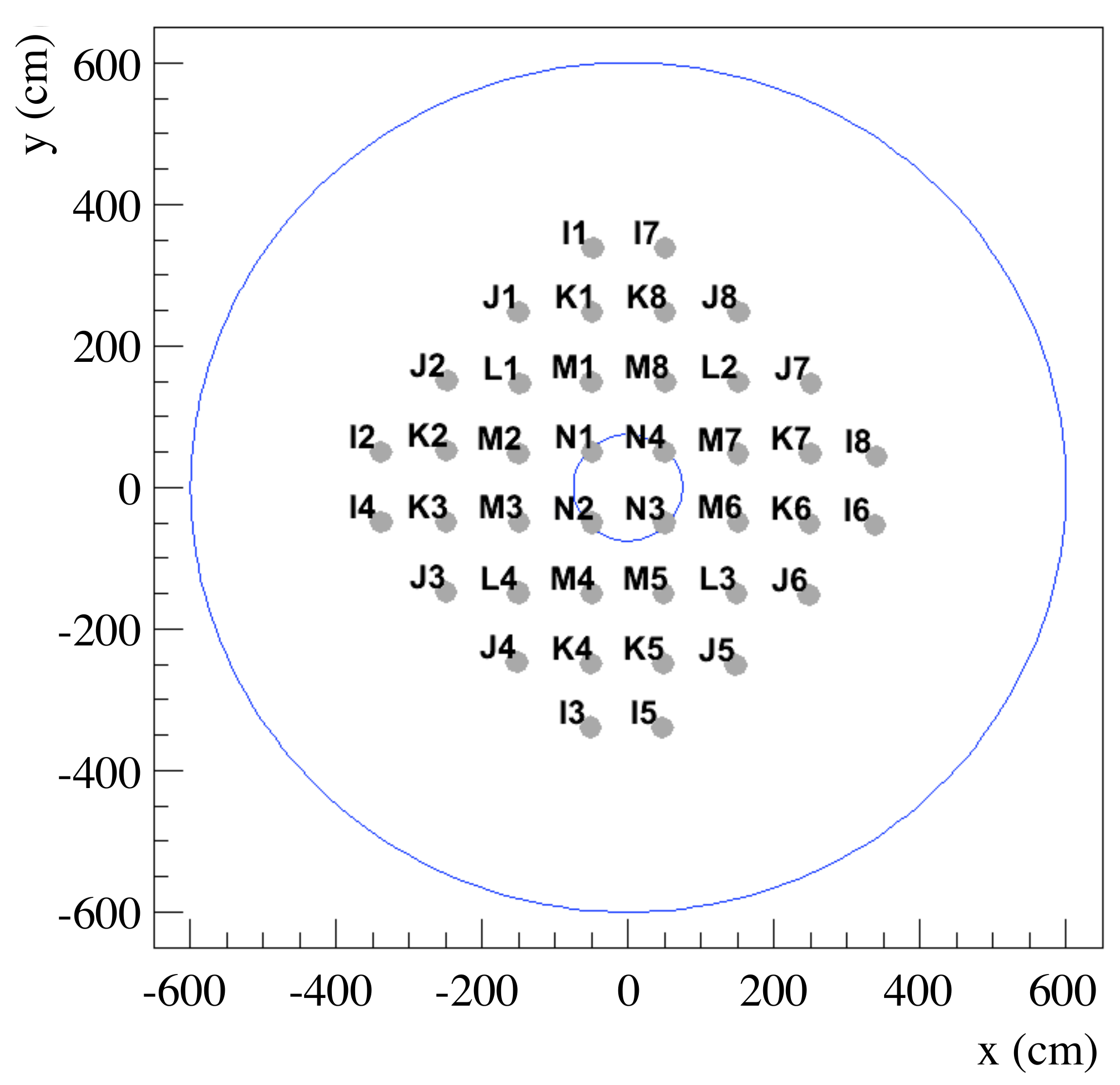}
\caption[NCD array]{The positions of the NCD strings projected onto the plane of the AV equator.  The array was anchored on a square lattice, but had cylindrical symmetry.  The letters denote strings of the same length and distance from the center of the AV.  Strings I2, I3, I6 and I7 were filled with $^4$He.  The outer circle is the AV equator and the inner circle is the neck of the AV through which the NCDs were deployed.  The NCD string markers are not drawn to scale.}
\label{array}
\end{center}
\bigskip
\end{figure}

The NCDs met other physical constraints.  The array needed to be large enough to detect a significant number of the neutral-current disintegrations without blocking a substantial amount of light to the PMT array surrounding the D$_2$O.  The NCDs also had to survive in ultrapure water at absolute pressures up to 3.2 atmospheres (atm) and at $10^\circ$C for several years.  Furthermore, the NCDs could not exceed the 370-cm maximum length set by the size of the mine hoist cage used to bring material to the SNO site.  In order to deploy vertical detectors with lengths between 9 and 11 m, smaller individual NCDs with lengths of between 200 and 300 cm were constructed, which were then joined together to form detector `strings' with electrical connections and water-tight seals.  A schematic of a single NCD string is shown in Figure \ref{ncd}.  A total of 36 NCD strings filled with $^3$He-CF$_4$ were deployed in the SNO detector, along with four NCD strings filled with $^4$He-CF$_4$ to study non-neutron backgrounds.  Figure \ref{array} shows the location and names of the forty NCD strings in the 12-m diameter SNO AV.  Each string was anchored with an acrylic anchor ball to an acrylic socket on the bottom of the AV on a square lattice with a nominal 1-m spacing.  The buoyancy of the NCDs and cables made the arrangement of the array vertical to better than 1$^{\circ}$, which has been confirmed with studies of the NCD shadows in the PMTs.  

\section{Design and construction of the NCD array}
\label{Const}

\subsection{Overview}

The bodies of the NCD proportional counters were made of nickel that was purified and formed into 5.08-cm-diameter, 2-meter-long tubes by means of a chemical-vapor-deposition process.  After the nickel tubes were received from the manufacturer, they were cut to length, weighed, leak-tested, straightened, flared at one end, electropolished, acid etched, rinsed in alcohol, and baked under high vacuum overnight.  The tubes were then assembled into 2- to 3-meter-long proportional counters by laser-welding sections together, adding electrical feedthroughs and endcaps at either end, stringing the anode wire, and filling them with the purified gas mixture.  The majority of this NCD construction work was performed in a Class-100 clean room at the University of Washington in Seattle.  Small parts were prepared in a clean room at Los Alamos National Laboratory.  Clean-room fabrication was essential because low particulate levels inside the finished NCDs reduce both radioactivity and the generation of spurious pulses.

Three or four NCDs were connected together with electrical couplers to form a detector string with a single readout cable.  One end of each NCD is slightly flared so it can slip over the straight end of another NCD, allowing structural attachment and welding of the NCDs into strings.  A cable assembly at the top and a delay line at the bottom complete each string.  The welding of NCDs, cables, and delay lines into complete strings occurred before and during deployment of the NCD array into the SNO detector. 

Physical parameters of the NCDs are shown in Table \ref{Mechanical}.  The diameter of the NCD tubes was minimized to reduce the obstruction of Cherenkov light in the PMT array, while still providing good neutron-capture efficiency.  A smaller NCD diameter increases the wall effect, although gas pressure and composition are also important factors in controlling this effect.  The minimum thickness of the nickel walls was driven by the requirement that the tubes not collapse when evacuated prior to gas fill, which, by choice of gas pressure, provided a more rigorous constraint than the pressure differential when the NCDs are submerged in SNO's heavy water.  Details of the NCD design and materials selection are given in the following Sections.  Additional information about the NCD design has been published in several dissertations and theses \cite{Shorty, Thornewell, MCBrowne, Heeger, Miles, Stonehill, Duba}.

\setlength{\tabcolsep}{0.03in}
\begin{table}[htb]
\bigskip
\caption{Physical parameters of the NCDs}
\label{Mechanical}
\medskip
\begin{center}
\begin{tabular}{ll}
\hline
Outer diameter & 5.08 cm \\
Wall thickness (nominal) & 370 $\mu$m \\
Wall thickness (measured) & 305 -- 533 $\mu$m \\
Lengths & 200, 227, 250, 272, 300 cm \\
Anode wire diameter & 50 $\mu$m  \\
Gas pressure & $2.5 \times 10^5$ Pa ($2.50 \pm 0.01$ atm)\\
Gas mix (by pressure) & 85\%:15\% $^3$He:CF$_4$ or $^4$He:CF$_4$\\
Weight  & 525 g/m \\
\hline
\end{tabular}
\end{center}
\bigskip
\end{table}

\subsection{Tube bodies}
\label{tubes}

Most organic materials (plastics, polymers, etc.\/) have very low thorium and uranium contamination levels. However, the fact that helium permeates such amorphous materials suggests the use of metal or metal-coated walls for proportional counter bodies instead. Technical difficulties associated with metal-coating materials such as acrylic led to the choice of metal walls.  Another advantage of metals is their high yield strength, which allows the total amount of material to be reduced by using thinner walls. Based on these considerations, the NCD bodies were made from ultrapure chemical-vapor-deposited (CVD) nickel.  Nickel was chosen for its strength, its chemical inertness in ultrapure water, and its ability to participate in the chemical reactions involved in the CVD process.  Under pressure, at around $50^\circ$C, nickel combines with carbon monoxide to make gaseous nickel carbonyl (Ni(CO)$_4$).  Upon reaching $175^\circ$C, nickel carbonyl decomposes again into nickel and carbon monoxide.  The nickel was deposited onto an aluminum mandrel to make the tubes for the NCD bodies.  This CVD process allowed the nickel to be separated from impurities, particularly thorium and uranium, which can form carbonyls \cite{carb1, carb2} but not reversibly at these temperatures.  Most metals in the Earth's crust contain on the order of 1 -- 10 $\mu$g of $^{232}$Th and $^{238}$U per gram of material.  By using the CVD process, the $^{232}$Th and $^{238}$U content of the NCD nickel was reduced by six orders of magnitude.  The CVD-nickel tubes used for the NCDs were produced by Mirotech, Inc. and by Chemical Vapour Deposition Systems, Inc.,\footnote{A subsidiary of Chemical Vapour Metal Refining, Inc., Toronto, Ontario, Canada, URL: www.cvmr.ca.} a company that subsequently acquired Mirotech's assets.  Some properties of the CVD nickel are shown in Table \ref{Properties}.

\setlength{\tabcolsep}{0.1in}
\begin{table}[htbp]
\bigskip
\caption{Properties of CVD nickel (courtesy Mirotech, Inc.)}
\label{Properties}
\medskip
\begin{center}
\begin{tabular}{ll}
\hline
Specific gravity & 8.871 \\
Yield strength $||$  & 440 MPa \\
Yield strength $\perp$  & 600 MPa \\
Ultimate tensile strength & $\sim$ 640 MPa \\
Elongation & $\sim$ 25\% \\
Modulus of elasticity & 178 GPa  \\
Residual stress (surface) & 30--60 MPa tensile \\
Coefficient of thermal expansion & $13.1 \times 10^{-6}$ K$^{-1}$ \\
Thermal conductivity & 88 W m$^{-1}$ K$^{-1}$ \\ 
\hline
\end{tabular}
\end{center}
\bigskip
\end{table}

Once pure materials are obtained, great care must be taken to prevent contamination. One concern is cosmogenic activity induced in otherwise pure materials while they are stored on the Earth's surface.  During the construction of the NCDs, the nickel tubes were manufactured faster than the NCDs were assembled, so it was desirable to store the tubes in an underground location.  An adit used to test tunnel-boring equipment for the Superconducting SuperCollider was identified near Index, Washington, less than two hours from Seattle.  When the first group of tubes was brought back to Seattle from this storage location, it was found that they had high rates of 5.3-MeV alpha events from $^{210}$Po decay.  The nylon bags in which the NCDs were stored at Index were not sealed, and subsequent measurements in the adit indicated a very high radon level of about 900 pCi/L \cite{MCBrowne}.  Radon daughters such as $^{210}$Pb and $^{210}$Po had plated onto the inner and outer surfaces of the nickel tubes.  In NCDs constructed from these tubes, the event rate was about 10$^5$ events/m$^2\cdot$day around the 5.3-MeV peak (the deployed NCD array has a surface area of 63.5 m$^2$), with several thousand events per day falling into the energy region of the neutron-capture signal \cite{MCBrowne}.  

This rate was unacceptably high, so subsequent tubes were electropolished to remove plated radon daughters from the surface of the nickel.  This electropolish was in addition to an acid etching procedure that had been developed to remove other contaminants such as aluminum oxide left over from the mandrel.  Electropolishing was necessary to remove polonium that could not be removed by acid etching because polonium displaces nickel into acidic solution and thus redeposits on a freshly etched surface.  The earliest set of NCDs constructed were built prior to storage at Index and prior to the development of the electropolishing procedures, thus these tubes were only acid etched.  The remaining tubes underwent both the electropolish and the acid etch.  Additionally, about half of the tubes had been stored at Index and required an acetic acid pre-wash to remove $^{210}$Pb prior to the standard cleaning processes.

The electropolishing was carried out with a custom apparatus in which reagents were transferred from storage to use by means of air pressure.   The interior of the tube was polished by inserting a 2.5-cm-diameter nickel cathode along the axis of an NCD and filling the interspace with 1 M sulfuric acid. A current of 620 A was passed between the NCD and the cathode.  A good finish and removal of polonium were obtained once the temperature of the solution had reached $55^\circ$C.  Two electropolish cycles, removing a total of 20 $\mu$m of nickel, were required to bring the inside surfaces to acceptable levels of $^{210}$Po activity.  The exterior surfaces were lightly electropolished to remove 2 $\mu$m of nickel in a separate apparatus in which the cathode was a stainless steel tube surrounding the NCD.  Alpha activity on the outer surfaces is a source of neutron background via the $^{17,18}$O($\alpha$,n) reaction, but only at relatively high alpha rates \cite{salt2}, so only one electropolish cycle was required for the outside surfaces.  The procedures needed to assure removal of polonium were assessed with the aid of contaminated test sections of CVD nickel and a low-background proportional counter.

All the tubes were acid etched at room temperature in a solution of 0.7\% hydrofluoric and 25\% nitric acid for 10 minutes, then transferred to a 5\% nitric acid solution for one minute, and finally to two weak ($10^{-4}$ M) acetic acid rinses for one minute each.  The acid etch removed 1 $\mu$m of nickel from each surface of the tubes.  Thus a total of 24 $\mu$m of nickel was removed from the tubes, which had originally been about 0.4 mm thick.  After the acid etch, the tubes were rinsed in deionized water, then transported immediately into the cleanroom, placed into a vacuum baking apparatus, pumped to $10^{-4}$ Pa with a turbomolecular pump, and baked for 12 hours at 150$^\circ$C. On removal from the chamber, the tubes were placed in sealed nylon film sleeves to reduce contamination by subsequent handling.

\subsection{Endcaps and electrical feedthroughs}
\label{endcapSection}

\begin{figure}[htbp]
\begin{center}
\includegraphics*[width=\columnwidth]{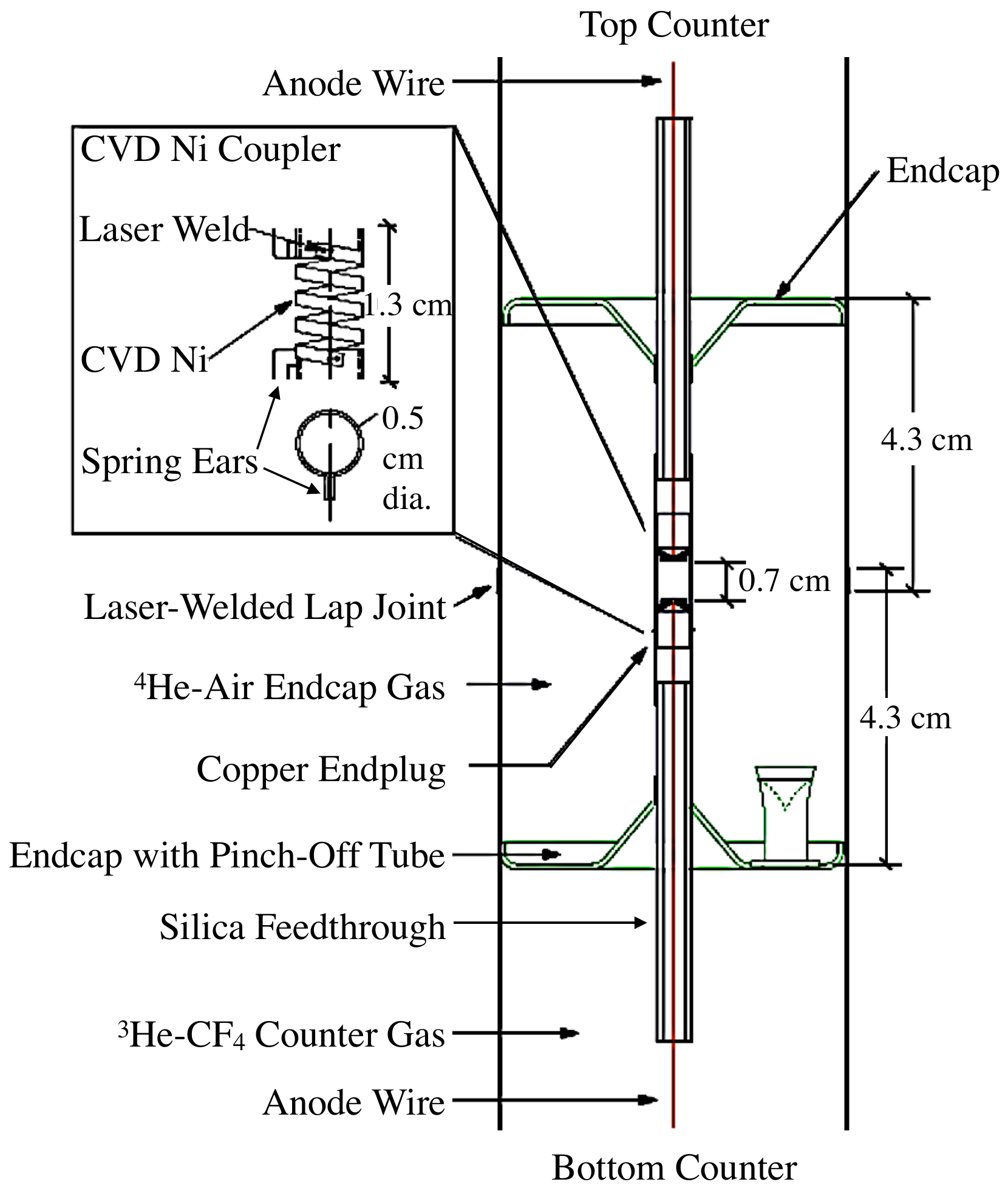}
\caption[NCD endcaps]{The endcap regions of two joined NCDs.  A CVD-nickel coupler was used to connect the copper tips of the two silica feedthroughs to make the electrical connection between the NCDs.}
\label{fig:endcap}
\end{center}
\bigskip
\end{figure}

Once the cleaned nickel tubes were cut and welded to the appropriate length, endcaps were welded into the ends of each NCD.  The endcaps were positioned 4.3 cm from each end of the nickel tube bodies to provide space to make electrical connections between individual NCDs.  The body of each endcap was made of CVD nickel formed on a stainless steel mandrel, then acid etched using a similar process to that used for the nickel tubes.  Each endcap contains a Heraeus-Amersil Suprasil T-21 fused-silica high-voltage feedthrough tube that insulates the anode wire from the nickel bodies of the NCDs.  The silica feedthrough extends between 2.5 and 2.8 cm into the NCD, producing a multiplication-free region at either end of the NCD to reduce the effects of the locally-distorted electric field.  The silica feedthrough also extends out of the endcap face 3.9 cm to the copper endplug, allowing electrical contact between NCD anode wires to be made at the endplugs using in-house fabricated CVD-nickel spring couplers.  The endcap region of two connected NCDs is shown in Figure \ref{fig:endcap}.

Techniques developed in collaboration with I.~J. Research, Inc.\footnote{I. J. Research, Inc., Santa Ana, CA, URL: www.ijresearch.com.} were used to metallize the silica feedthroughs and solder them with high-purity 96.5:3.5 eutectic SnAg alloy solder from Indium Corp. of America.\footnote{Indium Corporation of America, Utica, NY, URL: www.indium.com.}  The silica tubes were specially prepared to allow them to be soldered into the nickel endcaps.  The inside was coated with pyrolytic graphite and bands of chromium and nickel were put down by argon-ion sputtering on the outside. The chromium provided good adherence to the silica.  For soldering nickel, the surfaces were prepared with Acid Flux Number 4 from Indium Corp., and for soldering copper, the surfaces were prepared with technical-grade abietic acid.  The endplug into which the anode wire was soldered is made of an oxygen-free high-conductivity copper alloy, C10100.  

Making a seal between nickel and silica is difficult due to the mismatch in thermal-expansion coefficients. As the solder joints cool, the nickel contracts more than the silica, compressing the joints.  The nickel at the joints was made very thin (0.2~mm) so that it yields without breaking the silica, which is extremely strong under compression.  The endcaps were formed with a conical section to improve compliance under temperature variations.  The resultant joints are nevertheless sensitive to temperature fluctuations, so care was taken not to expose the finished NCDs to temperature extremes (below freezing or above about $40^{\circ}$C) during shipping and installation.  

The fact that helium permeates through glassy materials such as fused silica was a concern in the design of the endcaps.  Four short prototype NCDs containing the standard $^3$He-CF$_4$ gas mix and standard feedthroughs were placed in a whole-body leak test chamber at room temperature.  The detectors had been filled five weeks previously.  They showed $^3$He leak rates of $0.9 \times 10^{-7}$ to $1.3 \times 10^{-7}\ \mbox{STP-cm}^3$/s.  The endcaps were subsequently tested individually to confirm that each gave half the rate of the whole detector, and that the pinch-off seals, which seal the gas fill tubes, were leak tight.  At a permeation rate of $0.6 \times 10^{-7}\ \mbox{STP-cm}^3$/s per endcap, the loss rate is 0.6 STP-L/yr for the 312 deployed endcaps, which is 0.04\% of the total inventory of 1715 STP-L in the 398 m of deployed detector.  This permeation rate is within the initial specification of a gas loss of less than 0.1\% per year.  Simulations indicate that helium permeation into the endcap regions at this measured rate would lead to a 0.01\% change in the overall neutron-capture efficiency of the array over the two-year time period of the NCD phase.  Hence there are no major concerns with gas permeation through the feedthroughs at the measured levels.

\subsection{Anode wire}
\label{wireInfo}

Each NCD was strung with a 50-$\mu$m diameter hard-drawn copper anode wire made by California Fine Wire.\footnote{California Fine Wire Company, Grover Beach, CA, URL: www.calfinewire.com.}  The NCDs were strung in the vertical orientation by weighting the end of a length of wire with a needle.  Physical parameters of the anode wire are collected in Table~\ref{wire}. Copper is an excellent material for the anode, since it has good electrical properties and can be purified to very low levels of thorium and uranium, less than 10 pg per gram \cite{HamishSTR}.  It was unsuitable for the NCD bodies because of the softness and porosity of electrodeposited copper, and due to the amounts of radioactive lead introduced with the required mass of copper.  CVD nickel was unsuitable for the anode wire because nickel is ferromagnetic and exhibits high skin-effect losses.

\setlength{\tabcolsep}{0.1in}
\begin{table}[htb]
\bigskip
\caption{Anode wire summary}
\label{wire}
\medskip
\begin{center}
\begin{tabular}{ll}
\hline
Material & Copper 10100, hard drawn \\
Diameter & 50 $\mu$m \\
0.2\% yield strength & 380 MPa \\
Tension & 150 MPa \\
Critical stability length & 7.3 m \\
Resistance & 8 $\Omega$/m \\
\hline
\end{tabular}
\end{center}
\bigskip
\end{table}

The chosen anode wire thickness balances several factors.   A thicker wire decreases space-charge effects, but requires higher voltage to achieve the same gas gain.  Higher voltage is problematic because the rate of microdischarge events increases with operating voltage.  A thinner wire provides increased gas gain for a given voltage, but has a higher resistance per unit length, increasing resistive signal losses.

Annealed copper does not have sufficient strength to tolerate the stresses placed on it during tensioning, soldering, and detector cooling operations, so hard-drawn copper wire with a yield strength of 380~MPa was used.  In addition, hard-drawn copper has good creep properties and therefore will not sag with time.  In tensioning the wire, the total stress was limited to less than 190 MPa, a factor of two below the yield stress.  The mechanical stresses arising from cooling the wire after soldering and when inserting the NCDs into the D$_2$O at $10^\circ$C were about 30 MPa.  After accounting for this static stress from cooling, 160 MPa was the maximum stress that can be applied for tensioning the wire, so a tension of 150 MPa was used for the NCD anode wire.  During horizontal testing prior to deployment, catenary droop caused a position-dependent gain shift due to non-uniform displacement of the wire from the center of the detector.  At a tension of 150 MPa the deflection was less than 0.08 cm, causing a gain shift of less than 1\%.  This gain shift was eliminated when the NCDs were oriented vertically after deployment.

Another mechanical consideration was the uniformity of the wire diameter.  Diameter changes in the wire cause gain shifts in the detector of approximately 5\% for a 1\% variation in wire diameter.  For a 50-$\mu$m wire, manufacturers give a tolerance of $\pm 0.6$~$\mu$m or about 1\%.  However, experimental measurements indicated the position-dependent gain shift over a 3-m NCD to be around 1\%, which implies that the wire uniformity is better than 0.2\%.

\subsection{Gas fill}
\label{gasfill}

\begin{figure}[htb]
\begin{center}
\includegraphics*[width=\columnwidth]{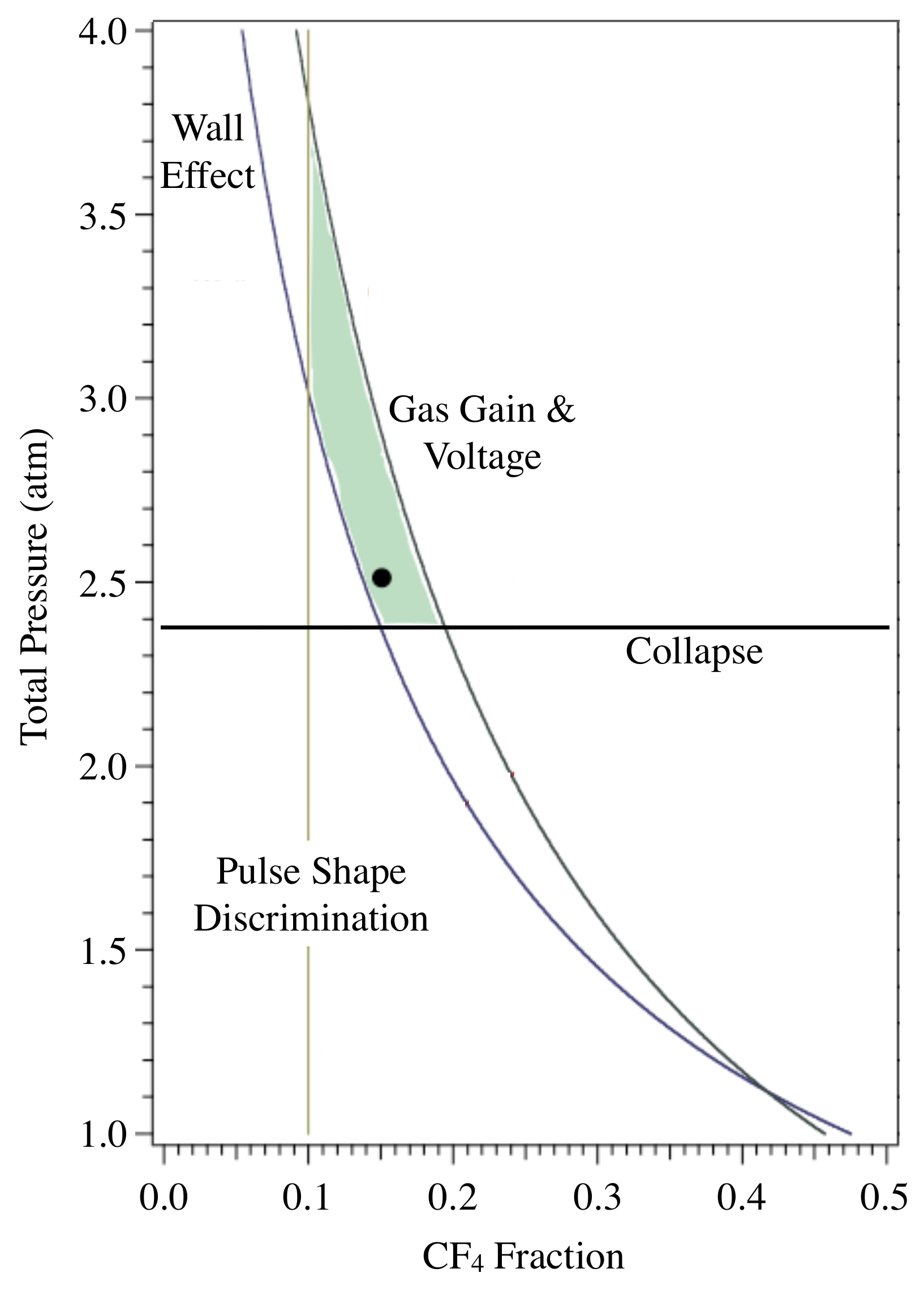}
\caption[gasfrac]{Optimization of gas fill fraction and pressure.  The shaded region shows the CF$_4$ fraction and total pressure that fit the scientific and mechanical constraints (explained in the text).  The dot shows the pressure and CF$_4$ fraction chosen for the NCD array.}
\label{gasfrac}
\end{center}
\bigskip 
\end{figure}

At the upper end of each NCD, the endcap has an oxygen-free high-conductivity copper (C10100) tube through which the NCD was filled with $2.50 \pm 0.01$~atm of gas.  The gas is a 85:15 mixture (by pressure) of helium and carbon tetrafluoride (CF$_4$) quenching gas, with a $^3$He density of $1.07 \times 10^{-4}$~g/cm$^3$.  For background studies, $^4$He replaces $^3$He in 16 of the 156 deployed NCDs, since $^4$He does not capture neutrons.  The gas mixture and total pressure were optimized based on several factors, as shown in Figure \ref{gasfrac}.  The tightest constraint on the CF$_4$ fraction came from the competing requirements that the wall effect be minimized, which suggests a larger CF$_4$ fraction, and that the operating voltage be minimized without sacrificing gas gain, which suggests a smaller CF$_4$ fraction.  These two competing effects are also somewhat dependent on the total pressure.  Another constraint on the CF$_4$ fraction arose from the fact that the track lengths of alphas and proton-triton pairs are comparable to the NCD diameter for CF$_4$ fractions below about 10\%, which negatively impacts the ability to perform pulse-shape discrimination to separate neutrons from alpha backgrounds.  In addition, CF$_4$ is critical for increasing the drift speed, which improves the signal-to-noise ratio and pulse-shape discrimination capabilities.  The optimal total pressure was primarily driven by the fact that below a gas pressure of about 2.4 atm, the tubes would collapse under the water pressure at the bottom of the SNO AV unless the nickel walls were made thicker, increasing radioactive backgrounds.  The neutron-capture efficiency (not shown in Figure \ref{gasfrac}) was also a factor in this optimization, although the cross-section is large enough that this factor does not introduce additional constraints.

Gas purification was very important to remove electronegative impurities (e.g. O$_2$) and to reduce backgrounds from tritium contamination.  The $^3$He used for the NCDs was obtained from Los Alamos National Laboratory (originally from Westinghouse Savannah River Company\footnote{Now Washington Savannah River Company, LLC, a subsidiary of Washington Group International, Boise, ID, URL: www.srs.gov and www.wgint.com.}).  The $^3$He was produced by tritium decay and, prior to purification, had about 1 mCi/STP-L of tritium in the form of water vapor (HTO) and some free hydrogen (HT).  This amount would produce $4.4 \times 10^{8}$ decays/s in a 2-meter NCD.  While the beta-decay of tritium deposits 6 keV in the gas on average, pile-up of tritium decays could produce events with enough energy to cause a background to the neutron-capture signal.  The $^3$He was purified to reduce the tritium level below 2.7 nCi/STP-L.  This limit represents a 1\% probability that a tritium decay occur in a 10-$\mu$s event integration time.  The NCD waveforms were digitized for 15 $\mu$s, but this limit is adequate to make pile-up a negligible background.  

Maintaining consistent gas mixture and pressure was crucial to achieving uniform gas gain throughout the NCD array.  A dedicated gas-handling system designed and constructed at Los Alamos National Laboratory was used for this, as well as to reduce the tritium contamination of the $^3$He.  Up to seven NCDs could be filled at once on a manifold.  For each set of NCDs, the appropriate amount of $^3$He was introduced into holding tanks, after passing through a single-pass cold trap with a charcoal sieve, a SAES\footnote{SAES Getters, URL: www.saesgetters.com.} getter (model PS2GC50R1), and a cooled charcoal trap.  The $^3$He was then circulated through the SAES getter and the charcoal trap for a minimum of three hours.  The proper amount of CF$_4$ was added to the holding tanks and the mixture was circulated through another SAES getter (model PF33CF) for a minimum of 15 minutes.  This purification removed water vapor and free hydrogen, which are sources of tritium, as well as air and other contaminants.  The purified gas mixture was then introduced into the NCDs.  

After gas fill, but before the copper fill tube was sealed off, a low-energy spectrum was obtained to measure the tritium contamination in the gas.  Data were taken from an NCD for 10 minutes at 2225~V to verify that the tritium content of the $^3$He was below 2.7 nCi/STP-L.  In addition, an energy spectrum was taken with a $^{252}$Cf neutron source to verify the gas gain of the filled NCDs.  Data were taken at 1600 V for 10 minutes from all the NCDs that had just been filled, as well as from a standard NCD with known gain.  The positions of the neutron peaks on each NCD were recorded, then the NCDs were switched systematically to different electronics channels and another spectrum was taken.  By comparing the newly filled NCDs to the standard NCD and by comparing the data taken on the different electronics channels, it was possible to remove the effects of the electronics and determine the gas gain of each NCD relative to the standard.  If the neutron peak position of an NCD was within 3\% of the standard and its resolution was better than 5\%, it passed the neutron-source test.  

The $^4$He NCDs were treated differently because it was undesirable to contaminate the holding tanks with $^4$He and because tritium removal was unnecessary.  The $^4$He was mixed with the CF$_4$ in the NCDs, rather than in the holding tanks.  The CF$_4$ was passed through the CF$_4$ SAES getter on its way to the NCDs.  The $^4$He was not purified, since 99.9999\% pure $^4$He was used.  Because neutrons are not captured in the $^4$He NCDs, the gain and resolution were measured with a $^{241}$Am gamma source.  The 59.45-keV gamma produces an electron continuum with a dominant K edge from nickel at 51 keV.  Data were taken for 10 minutes at 1600 V with the amplifier gain raised by a factor of five compared to the $^3$He NCD gas test.  If the position of the 51-keV K edge indicated the correct gas gain, and the resolution was better than 16\%, the NCD passed.  

Any leak in the seal would not only allow gas to escape, affecting the gain of the detector, but would also allow impurities to enter the NCD.  Electronegative impurities absorb electrons from the ionization tracks, thereby degrading the signal strength and the resolution. In order for the detectors to work properly, such impurities must be kept well below the $10^{-6}$ level. Another adverse effect of a leak in the seal is that the `dead' volume between the individual NCDs may become filled with $^3$He, capturing neutrons that cannot be detected by the NCD and thereby making calibration of the efficiency difficult.  A mechanical crimp was used to seal and cut off the copper fill tube.  After being sealed, NCDs were stored in nylon bags to prevent contamination and packed for shipment to SNO, where they were monitored periodically prior to deployment to verify the stability of the neutron-capture peak position.

\subsection{Delay line and anchors}

Position information is desired to reconstruct the distribution of events from NC interactions and from AV backgrounds, as well as to establish connections between Cherenkov events and related neutrons.  The association of Cherenkov events with neutrons is particularly important in the detection of $\rm \overline{\nu}_{e}$ interactions from a supernova or, possibly, from the Sun.  The mean travel distance before capture for a neutron in D$_2$O is 113 cm, so position resolution requirements are not demanding.  Two-dimensional position coordinates were delivered by the 1-m grid layout of the NCD strings; the third (vertical) coordinate called for an electronic technique.  

A variety of position encoding techniques have been developed for gas counters \cite{Hendricks}. Charge-division and risetime encoding both depend on the use of resistive wires that function as diffusive delay lines, which is fundamentally incompatible with the need to extract time-development profiles of primary ionization.  The approach adopted for the NCD strings was to leave the remote end of the string unterminated (open) and to measure the time delay between current arriving directly at the preamplifier input and current reflected from the open end. Since the line is open, the low-frequency noise performance is determined by parallel noise alone, and is much improved.  Moreover, the full signal charge is collected from an open line, which further improves the signal-to-noise ratio.  The single-ended readout also reduced the amount of material in the heavy water, lessened interference with Cherenkov light, and minimized the number of cables running through the neck of the AV.

\begin{figure}[htb]
\begin{center}
\includegraphics*[width=\columnwidth]{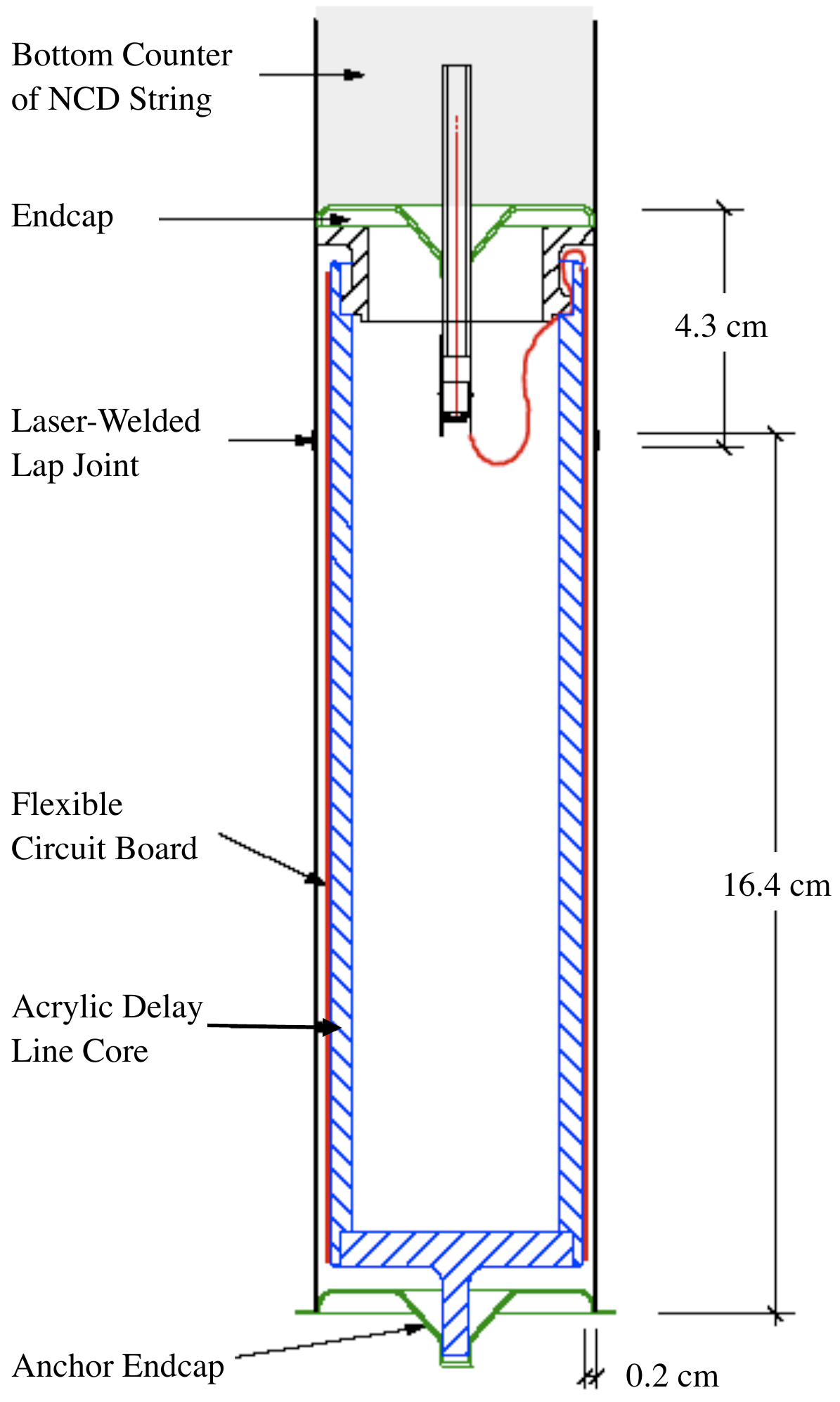}
\caption[NCD delay]{Schematic of the open-ended delay line attached to the bottom NCD of each string to provide position readout by pulse-reflection timing.}
\label{fig:delayline}
\end{center}
\bigskip
\end{figure}

The bottom of each string was terminated with an open-ended delay line, pictured in Figure \ref{fig:delayline}.  The delay line increased the effective length of the NCD string by about 45 ns each way, facilitating position readout for events that occurred near the bottom of the string by increasing the delay between the initial and reflected pulses.  The delay line consists of a flat meander circuit etched onto 0.05-mm copper-clad Kapton. No tinning or coating was used. The flexible circuit sheet was then wrapped around a 19-cm-long acrylic core, glued with a cyanoacrylate glue, and inserted into a CVD-nickel tube housing.  A pigtail from the circuit was soldered to a nickel clamp ring to make electrical connection to the copper tip of the feedthrough of the lowest NCD in a string.  This design gives the required delay and reduces high-voltage-induced microdischarges \cite{microDpaper}.  In-situ measurements determined the round-trip delay time to be $89 \pm 3$ ns, and the rise time to be about 7 ns.

At the bottom of the delay-line housing is an anchor endcap with four holes, to which Vectran braids were attached by means of CVD-nickel rings.  These Vectran lines come together and were strung through two acrylic anchor balls to form the anchor assembly.  The upper anchor ball was used only during deployment, while the lower anchor ball was inserted into an anchor point to attach the NCD string to the AV.  These anchor points are on the bottom of the AV on a nominal 1-meter grid with deviations as necessary to avoid bond lines in the AV, and their positions were surveyed to an accuracy of $\pm 3$ mm.  

\subsection{Cable and connectors}

The NCD cables provided the high voltage to the NCD strings, as well as coupled the signal from the anode wire to the electronics.  The NCD cable was rated for long-term underwater use without significant degradation of electrical properties. It has low levels of radioisotopes to avoid contributing a significant photodisintegration neutron background: less than 150 pg thorium per gram of cable (pg thorium/g) and less than 1750 pg uranium/g.  The cable is slightly buoyant in D$_2$O so that it conformed to the upper hemisphere of the AV.  

\begin{figure}[htbp]
\begin{center}
\includegraphics*[width=\columnwidth]{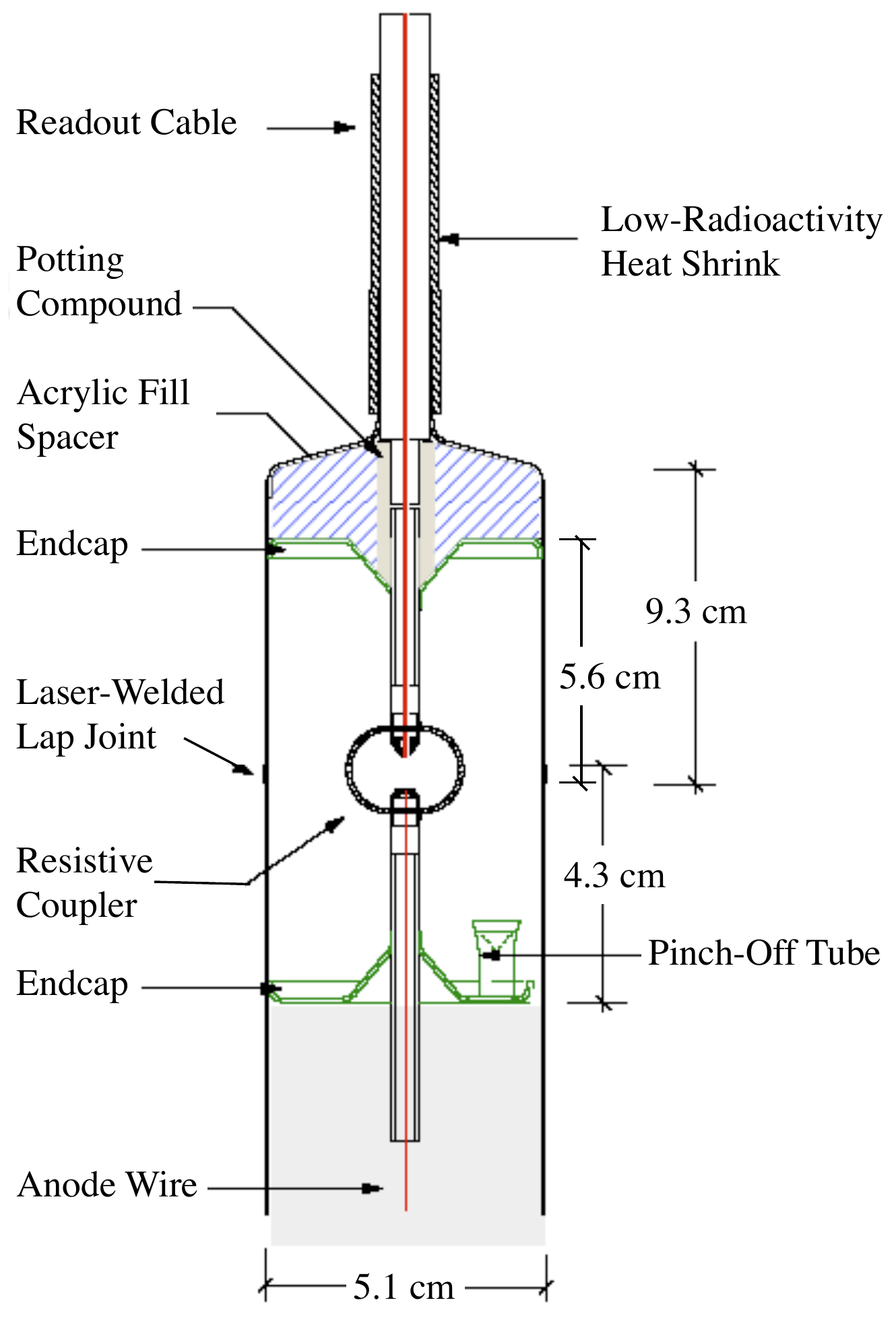}
\caption[NCD cable bell]{Schematic of the single-ended readout with the cable bell connected to the top NCD of a string.}
\label{cablebell}
\end{center}
\bigskip
\end{figure}

The cables range in length from 9 to 12 m and connected the NCD strings to the preamplifiers situated on the deck of the SNO cavity.  The cables were manufactured by South Bay Cable.\footnote{South Bay Cable Corporation, Idyllwild, CA, URL: www.southbaycable.com.}  The coaxial cable has a 0.5-mm-diameter central copper conductor, a 7.5-mm-diameter low-density polyethylene  core, a shield woven of $0.05 \times 1.0$-mm copper ribbon, and a high-density polyethylene (HDPE) foam jacket surrounded by a layer of HDPE serving as a water barrier.  The shield material is thin to improve buoyancy, so it cannot perfectly reject low-frequency noise because of skin effect penetration and resistance between the ribbons.  Characteristics of the NCD cables are given in Table \ref{cable_table} and the connection to the top NCD in a string is shown in Figure \ref{cablebell}.

\setlength{\tabcolsep}{0.1in}
\begin{table}[h]
\bigskip
\caption{Characteristics of the NCD cable}
\label{cable_table}
\medskip
\begin{center}
\begin{tabular}{ll}
\hline
Impedance & 93~$\Omega$ \\
Type & coaxial \\
Voltage rating& 3~kV\\
Outer diameter& $<0.890$~cm \\
Specific gravity& 0.995\\
Minimum bend radius & $7.5$~cm \\
\hline
\end{tabular}
\end{center}
\bigskip
\end{table}

The end of the cable connector that mates with the NCD string was made of a standard CVD-nickel endcap with a silica feedthrough.  The only modification required was that the hole in the copper tip be larger in diameter to accommodate the cable conductor.  To minimize reflections, the 93-$\Omega$ cable impedance is matched to the detector impedance of 415 $\Omega$ through a 325-$\Omega$ resistive coupler.  This coupler was made from a teflon cylinder with 0.3-$\mu$m Stablohm resistance wire from California Fine Wire strung back and forth around one half of the cylinder.  The wire is connected to CVD-nickel toothed rings in the teflon that slipped over the conducting tips of the silica feedthroughs in the cable connector and the uppermost NCD in the string.  The connection method proved to be less than completely reliable, as several strings developed poor contact at this point during the two years of operation.

The cable side of the connector was made from a `cable bell' formed by deposition of CVD nickel onto an aluminum mandrel that was subsequently etched away. A tube stub to which the cable was fitted extends from the cable bell.  Between the cable bell and the endcap is a piece of acrylic to provide low-radioactivity electrical insulation.  A good ground connection was provided by soldering the copper braid directly to the cable bell.  The assembly was potted with silicone through the silica and out to the copper braid.  After curing, part of the silicone was removed from the inside of the silica region to the shield of the cable because poor bonding to the silica was found to be a source of microdischarge.  The primary water barrier was provided by one layer of low-radioactivity heat shrink tubing and a thermosetting adhesive.  This tubing seals the jacket of the cable to the CVD tube stub protruding from the connector cable bell. The silicone provides a secondary water barrier.   The heat shrink tubing provides strain relief.

\section{Deployment of the NCD array}

\subsection{Cool-down phase}
\label{Stor}

After their construction in Seattle, the NCDs were transported to Sudbury and brought underground as quickly as possible to maximize the amount of time for cool-down of cosmogenic activity.  For the NCDs the only isotope of concern that produces a significant number of gammas above 2.2 MeV and has a half-life longer than 10 days is $^{56}$Co, with a 77-day half-life.  The $^{56}$Co is produced by spallation neutron interactions with $^{58}$Ni and $^{60}$Ni in the NCDs, and decays via electron capture and subsequent gamma emission.  Approximately 33\% of $^{56}$Co decays result in a gamma above the photodisintegration threshold of deuterium, thus $^{56}$Co constitutes a neutron-producing background in SNO. 

The nickel tubes comprising the NCD bodies were on surface long enough to essentially reach $^{56}$Co saturation (90\% saturation occurs after about 250 days).  If they had then been deployed into SNO with no underground cool-down time, five photodisintegration neutrons per day would be produced by $^{56}$Co decays \cite{MCBrowne}.  The final shipment of NCDs arrived underground in Sudbury in November 2002 and production data-taking in the NCD phase did not begin until November 2004.  The average NCD was underground for 22 half-lives of $^{56}$Co prior to the start of data-taking, reducing the neutron production rate from $^{56}$Co to approximately $10^{-6}$ neutron per day.

\subsection{Array optimization}

Since the solution to the Solar Neutrino Problem was not known at the time of the initial NCD design, the original size of the NCD array was conservatively based on the solar neutrino flux deduced from previous experiments, rather than the full Standard Solar Model flux.  Once the first results from SNO confirmed the SSM-predicted total solar neutrino flux, a smaller, optimized array could be used in the NCD phase.  A smaller array would still make a significant measurement of the NC reaction rate in a reasonable amount of time and would block less of the Cherenkov light produced in the CC reaction, allowing for a better measurement of the neutrino energy spectrum.  The original NCD array was designed to consist of 96 strings, ranging from 4 m to 11 m in length, for a total of 770 m.  Each string would have contained two, three, or four individual NCDs, so the original array would have consisted of 300 NCDs.  The array deployed for the NCD phase consisted of the 40 central strings from the original array, ranging from 9 m to 11 m in length.  The total length of the deployed array was 398 m, and it consisted of 156 individual NCDs.  

The final NCD array length and configuration were determined by maximizing the neutron-capture efficiency while minimizing interference with Cherenkov light produced by CS and ES reactions.  The optimal length was determined to be 40\% -- 50\% of the original array, with the CC spectrum determination and other systematics-limited analyses favoring a smaller NCD array and the NC day-night measurement and other statistics-limited analyses favoring a larger array.  A central configuration was chosen to maximize the neutron-capture efficiency, minimize the amount of NCD cable required, and simplify neutron calibrations and the geometry of the Cherenkov light loss.  The primary disadvantage to the central configuration is the fact that the NCD array does not sample large radii uniformly.  The chosen configuration reduces the expected NC neutron-capture efficiency on the NCD array from 44\% with the original array to 26\%, while decreasing the PMT shadowing from 16\% to 9\% for a uniformly distributed Cherenkov light source.  

With the decision made to deploy a reduced NCD array of 40 strings, it was desirable to use the NCDs with the lowest individual levels of intrinsic radioactivity.  These were chosen by analyzing data taken during the cool-down phase while the NCDs were stored underground in the SNO control room.  The number of electronics channels was limited at this time, so data were taken between June and December 2002 in three data sets covering different subsets of the NCD array.  Almost all of the NCDs were represented in at least one of these data sets and only NCDs for which there was analyzable data were selected for deployment.  The criteria used in NCD selection included good gas gain and resolution and low levels of alpha activity, particularly from uranium- and thorium-chain decays as opposed to surface $^{210}$Po activity.  Once the NCDs were selected, their string assignments were specified by matching the gains as closely as possible.  The standard deviation of the gains for the NCDs within each string is smaller than 1\%.

\subsection{Predeployment activities}
\label{PreD}

Each NCD string consisted of three or four individual NCDs electrically connected and then welded together.  To weld together two NCDs, a $50 \mbox{-W}$ turnkey Lumonics\footnote{A subsidiary of GSI Group, Warwickshire, England, URL: www.gsiglasers.com.} 1024-nm Nd-YAG pulsed laser with 1.3 kW peak power was used.  One end of each NCD is flared slightly, so it could slip over the straight end of the adjacent NCD.  The beam from the laser was directed at this overlap and rotated relative to the NCDs, creating a series of craters melted into the nickel that fused the two NCDs together.  During welding, the NCDs were held in a stainless steel weld fixture that had been custom built at the University of Washington.  The weld fixture was hermetically sealed during welding to prevent the possibility of personnel injuries.  A HEPA filtration system on the weld fixture prevented nickel dust from contaminating the clean SNO laboratory.  To verify the integrity of the electrical connections, time domain reflectometry (TDR) measurements were made of the NCDs before and after they were connected and welded.  It was crucial to ensure that the welds were leak tight, so 10 cm$^3$ of $^4$He tracer gas was injected into the endcap region just prior to welding, allowing a series of three leak tests to be performed after each weld.    

NCD deployment was done in two stages to reduce the time that the SNO detector was offline and to minimize any potential for contamination of the D$_2$O.  During the predeployment welding stage, which occurred while SNO was still taking data in the salt phase, the NCDs were welded together horizontally into sections not exceeding 5.5 m.  Longer NCD sections could not be positioned vertically over the neck of the AV for deployment due to limited overhead clearance.  At this time, the delay lines and the high voltage/readout cables were also welded to the NCDs.  Of the 196 required welds, 148 could be done during predeployment welding, greatly reducing the SNO detector down time for deployment itself.  In addition, the predeployment welding stage allowed the welding procedures to be tested and refined before the critical deployment stage.  

Prior to storage in a clean nylon bag, the welded NCD section was cleaned with a custom-designed Teflon vacuum head that blew dry nitrogen gas onto the NCD while a clean-room vacuum removed any dust that was dislodged.  Data were taken from each welded section awaiting deployment to search for high voltage discharge or other problems.  If discharge was evident, the cause was often nickel particulates ejected from the inner NCD wall by the welding process.  This problem was due in large part to inadequate control over the laser power, which was aggravated by a repair of the laser during the predeployment welding.  In a few cases, the laser power was so high that the laser penetrated both layers of nickel, causing leaks in the endcap region that could allow D$_2$O to contact the high-voltage connections.  Because of these problems, about a dozen welds were repaired prior to the start of deployment.  During these repairs, thin CVD-nickel backing rings were introduced that added another layer of nickel to the weld region.  The backing rings greatly reduced the chance of leaks and of weld dust settling on the couplers or other high-voltage regions.  These refined welding procedures were adopted for the welds performed during deployment.

\subsection{NCD deployment}
\label{Hard}

\begin{figure*}[p]
\begin{center}
\includegraphics*[width=1.75\columnwidth]{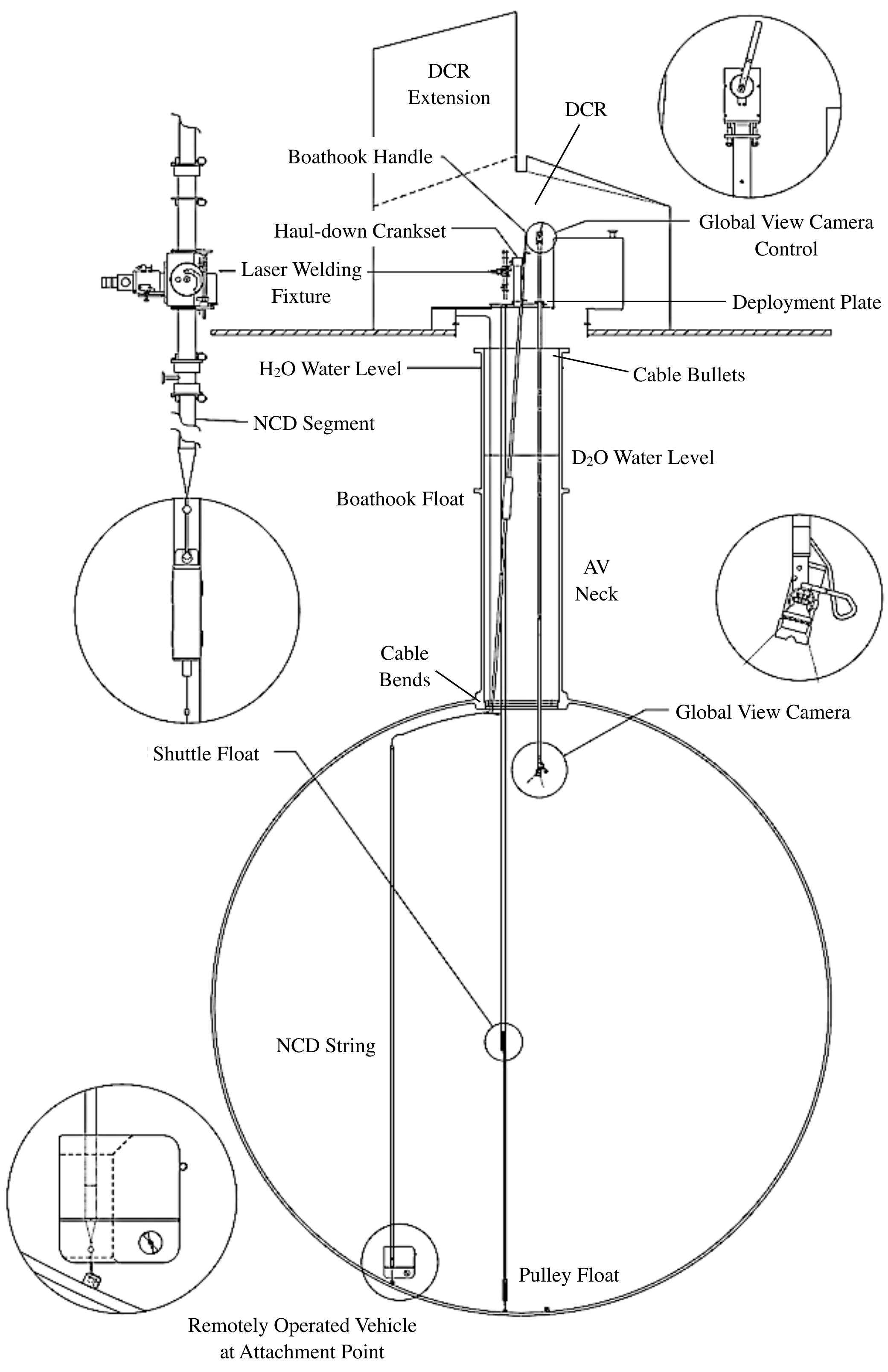}
\caption{The AV and DCR with the equipment used for NCD deployment.}
\label{fig:AVwHardware}
\end{center}
\bigskip
\end{figure*}

A significant amount of specialized equipment was required to deploy the NCD array into the SNO detector.  Figure \ref{fig:AVwHardware} shows the deck clean room (DCR) above the SNO detector, the AV, and the NCD deployment hardware. For the duration of deployment, this hardware was installed in the DCR or in the AV itself.  All deployment hardware was made of clean materials such as stainless steel or acrylic that are suitable for use in the DCR or insertion into the SNO detector.  Equipment that contacted the D$_2$O was tested prior to use to ensure that no contaminants could leach into the heavy water.   

A stainless steel deployment plate was installed over the neck of the AV.  The welding fixture was attached to the deployment plate in a vertical configuration on an arm that allowed it to swing out of the way when it was not in use.  There were four view ports on the deployment plate that could have gloves fitted to them if needed for manipulating NCD parts during deployment.  The NCD insertion port was at the bottom of a well about 0.5 m deep that allowed longer NCD sections to be deployed, reducing the number of welds required during deployment.  

To deploy an NCD string, the sections were first brought into the DCR after their nylon bags had been wiped down to remove contaminants that might have accumulated during storage.  A pulley attached to the ceiling of the DCR was used to position the sections vertically over the deployment plate.  At one end of the pulley cable was a cradle that attached to a clamp that could be placed around an NCD.  Counterweights attached to the other end of the cable provided upward force on the NCD section.  Once the nylon bag was removed and the NCD section was positioned vertically, the bottom end of the section was lowered into the insertion port.  Directly below the insertion port was a haul-down system consisting of two polypropylene floats on Vectran polymer fiber lines.  The lower float was anchored to an anchor point on the bottom of the AV.  The upper float, known as the shuttle float, was raised and lowered on the Vectran lines by turning the haul-down crank.  The lower anchor ball of the NCD anchor assembly was inserted into a socket on the shuttle float, remaining engaged in the socket as long as upward force (provided by the pulley counterweights or the buoyancy of the NCD section once in the heavy water) was applied to the anchor ball.    

Once the lower NCD section had been lowered to the welding position, the upper section was raised vertically using the pulley system and positioned in the weld fixture.  Two silicone cuffs were inflated to hold the NCD in place as the weld fixture was raised above the lower section and swung into the weld position above the insertion port.  The weld fixture was then lowered until the straight end at the bottom of the upper NCD section slid into the flared top of the lower section.  The two sections were welded together and leak tested, then a TDR measurement was made of the connected string to verify electrical continuity.  Once the final weld was verified, the NCD string was lowered further into the heavy water and the weld fixture was swung out of the way.  The cable bend and bullet used to engage the cable in the AV neck were installed on the cable.  Then the string was lowered to the bottom of the AV.  

A submersible remotely-operated vehicle (ROV) was used to maneuver the string into its anchor point. The ROV is a custom-built Phantom series ROV constructed by Deep Ocean Engineering, Inc.\footnote{Deep Ocean Engineering, San Leandro, CA, URL: www.deepocean.com.} to operate in ultrapure D$_2$O and manipulate NCD strings.  Specifications for this ROV are given in Table \ref{tabROV}.  Consideration was given during its design to SNO's stringent radioactivity limits on material leached into the heavy water.  Leach tests performed before deployment indicated the leaching of radium, but at acceptably low levels.  The equivalent masses of thorium and uranium that would leach into the D$_2$O during deployment were $2 \pm 1$~$\mu$g $^{232}$Th and $8 \pm 3$ $\mu$g $^{238}$U.  Additionally, assays of the D$_2$O performed while the ROV was in the heavy water and after its removal indicated no significant elevation of the $^{232}$Th and $^{238}$U levels in the D$_2$O.  Throughout this time period, the measured levels in the D$_2$O remained below the target level of 3.8~$\mu$g $^{232}$Th and about an order of magnitude below the target level of 30 $\mu$g $^{238}$U.

\setlength{\tabcolsep}{0.1in}
\begin{table}[htb]
\bigskip
\caption{Specifications for the ROV}
\label{tabROV}
\medskip
\begin{center}
\begin{tabular}{ll}
\hline
Weight & 55 kg \\
Displacement (tanks empty) & 55 kg \\
Trim weight & 32 kg\\
Dimensions: & 65.4 cm high\\
& 66.4 cm wide \\
& 59.1 cm deep \\
Propulsion: & \\
\hspace{0.5 in} 2 x horizontal & 60 W\\
\hspace{0.5 in} 2 x vertical & 60 W\\
\hspace{0.5 in} 2 x lateral & 60 W\\
Instrumentation: & Video camera\\
& 2 x 150 W lights\\
& Depth gauge \\
& Microphone \\
Umbilical: & 50 m long \\
& 1 cm diameter\\
\hline
\end{tabular}
\end{center}
\bigskip
\end{table}

To hold an NCD string, the ROV engages the upper anchor ball with an acrylic plate and socket mounted on the front of the ROV.  A camera on the ROV provides a close-up view of the socket and the anchor balls.  The ROV has two ballast tanks that could be purged with nitrogen gas or filled with up to 25.7 kg of water to make the ROV heavy enough to overcome the buoyancy of an NCD string, which varies by about 25\% depending on the string length.  The ROV is maneuvered by three pairs of thrusters controlled by two joysticks on a hand-held control unit.  The control unit also adjusts the trim of the ROV (the amount of automatic thrust provided by the vertical thrusters) and controls the solenoid valve that opens the ballast tanks.  The controls were connected to the ROV via a waterproof, neutrally-buoyant electrical umbilical bundle, to which the purge gas line is attached.

After engaging the upper anchor ball, the ROV operator flooded the ballast tanks to remove the NCD string from the shuttle float.  Then the ROV was parked on the floor of the AV near the haul-down, and neutron-source calibrations were performed using an $^{241}$AmBe neutron source encapsulated in the shuttle float.  The NCD readout cable was connected to a test preamplifier and high voltage was applied.  The shuttle float was raised a predetermined number of turns of the haul-down crank to place it near the middle of the lowest NCD in the string.  Data were taken for five minutes to verify that the NCD produced a good neutron-capture spectrum.  Then the shuttle float was raised to the center of the next NCD in the string and the data-taking was repeated.  In this way, each of the three or four individual NCDs in the string was verified.  The voltage was then turned off and the preamplifier was disconnected.  The $^4$He NCD strings were verified by using gammas from the $^{241}$AmBe source and confirming the location of the 51-keV K edge in the spectrum.

After the source tests, the ROV was `flown' to the NCD string's anchor point at a speed of a few meters per minute, assisted by a global-view camera mounted near the base of the AV neck.  A previously-mounted nickel nameplate on the anchor point was checked to verify its identity.  The ROV operator then inserted the NCD string's lower anchor ball into the anchor point.  Once the ball was engaged, the ROV's ballast tanks were purged to release its hold on the NCD string.  

The final step of deployment was securing the NCD cables to prevent them from becoming tangled in the manipulator lines used for positioning calibration sources in the SNO detector.  Each cable followed a curve of radius about 1 m from the top of the NCD string until it became tangent with the surface of the AV, then it followed the curve of the AV to the bottom of the neck.  A rigid polypropylene quarter-circle known as a cable bend attached the cable to an acrylic ring mounted just inside the bottom of the AV neck, forcing the cable to make a 90$^\circ$ bend and lie flat against the wall of the neck.  A long pole with pegs on the bottom was used to attach the cable bend to this acrylic ring.  A third camera provided views of the neck region to aid in this process.  Polypropylene cable bullets attached to each NCD cable slipped into indentations on a metal ring at the top of the neck, supporting and tensioning the cable in the neck to ensure no slack.  The dry end of each cable emerged sideways through a compression fitting to attach to the NCD preamplifier, located in a circular raceway above the top of the neck.  Finally, black heat shrink was applied over the compression fitting to ensure that no light would enter the AV.

\begin{figure}[htb]
\begin{center} 
\includegraphics*[width=\columnwidth]{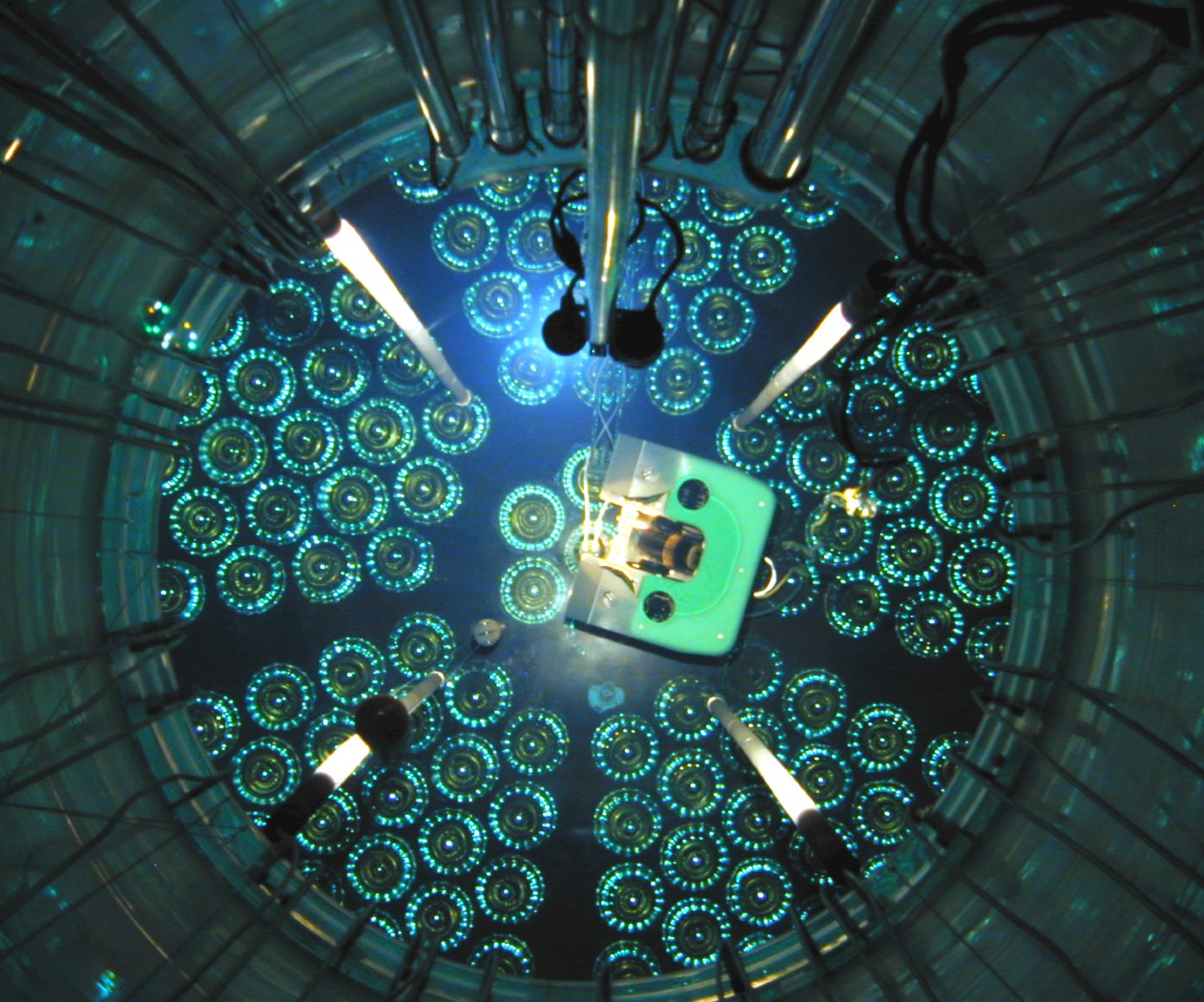}\\
\caption{A view down the AV neck after deployment of the NCD array, showing the ROV and four NCD strings, as well as the haul-down floats, the global-view camera, the NCD cables, and some water circulation pipes (at the top).  Some of the PMT array is also visible beyond the transparent AV.} 
\label{photos}
\end{center}
\bigskip
\end{figure}

NCD deployment began December 2, 2003 and the last string was deployed on February 12, 2004.  However, repairs to several strings were required, so the ROV was not removed from the heavy water until April 21, 2004.  After that, commissioning of the NCD array proceeded until November 2004.  Figure \ref{photos} shows the interior of the SNO detector after deployment of the NCDs.

\section{NCD electronics and data acquisition}

\subsection{Overview}

\begin{figure*}[htb]
\begin{center} 
\includegraphics[width=1.99\columnwidth]{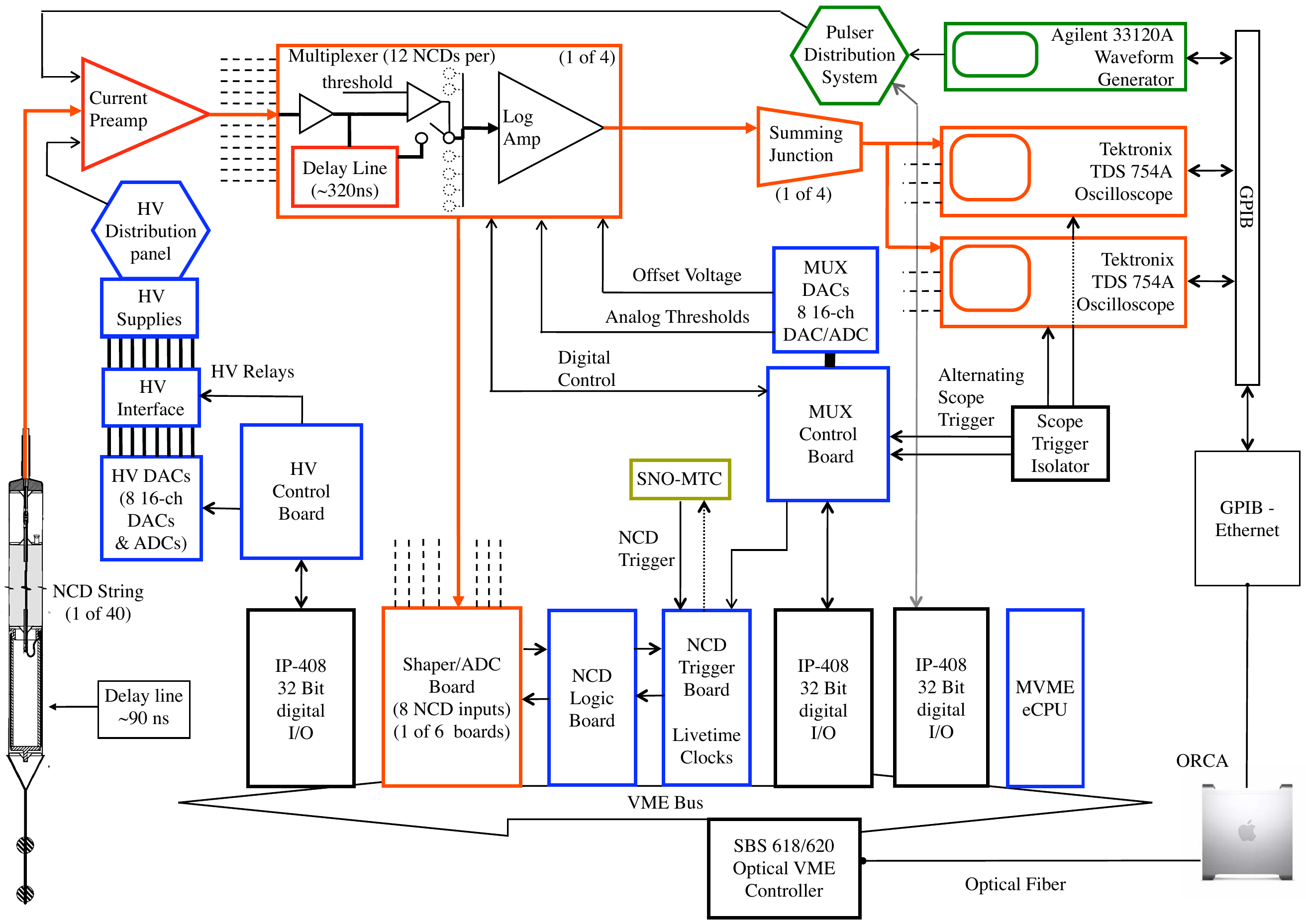}
\caption{Diagram of the NCD electronics systems.} 
\label{NCDelec}
\end{center}
\bigskip
\end{figure*}

The NCD electronics were designed with several goals: the ability to perform pulse-shape discrimination to distinguish neutron-capture signals from alpha backgrounds and spurious events, measurement of the total charge of the detected events, and collection of some information at the kHz event rates produced by a galactic supernova.  To satisfy these goals the NCD electronics have two independently-triggered readout systems: a fast data path using shaper-ADCs and a multiplexed digitizing path \cite{cox}.  Information about the digitized pulse shapes can be found in Section \ref{signals}.  Figure \ref{NCDelec} shows the basic schematic of the NCD electronics, which are briefly discussed in the following Sections.

\subsection{NCD preamplifier}

The NCD strings were cabled to current (trans\-resistance) preamplifiers designed at the University of Washington.  The NCD preamplifiers were located in a circular enclosure above the top of the AV neck in the DCR, minimizing their distance from the NCD strings.  The preamplifiers have two gain stages.  The input stage is a folded cascode configuration with negative feedback applied to the JFET gates.  The preamplifier utilizes four 2SK152 JFETs in parallel at its input and combines high transconductance with low input capacitance.  The input impedance is set to match the $93 \mbox{-}\Omega$ impedance of the NCD cable by a potentiometer that sets the current of the JFETs.  The first stage is buffered by an NPN emitter-follower that feeds a second gain stage.  The output buffer is an AD8055 operational amplifier.  The overall preamplifier design allows for a passband of 3 kHz to 45 MHz (limited to about 11 MHz by the multiplexer), which is adequate since lateral straggling widens the narrowest of NCD current pulses to approximately 33 ns.  The parallel and series noise generators at the input of the current preamplifier have been measured to be 1.5 pA~Hz$^{-1/2}$ and 0.41 nV Hz$^{-1/2}$, respectively, at 6 MHz.  The current preamplifier converts the signal to a voltage amplitude (1 $ \mu$A $= 27.5$ mV).  Additionally, the preamplifier served as the connection point for the high voltage and test pulser inputs.  The preamplifier's output cable was connected to a channel of the multiplexer located in one of the NCD electronics racks just outside the DCR.

\subsection{Multiplexer}

Signals from the preamplifiers entered two parallel buffer amplifiers, one that drove a cable to the shaper-ADC system that resided in a VME crate, and the other that drove a delay line and discriminator for use by the digitizing system (two oscilloscopes).  The delay line is a non-inductively wound RG-58 cable on a PVC spool, providing a delay of approximately 320 ns.  Droop caused by the frequency-dependent attenuation in the delay line is compensated by a peaking network.  This delay allows enough time to trigger the digitizing system without significant loss of event information.  The negative-going input signal and a $+1$ to $-40$ mV DAC set level are fed to an LT1016 comparator acting as a discriminator. Signals exceeding the discriminator threshold are sent to an AD8307 logarithmic amplifier. At threshold levels more negative than $-35$ mV, however, an LT1011 comparator sets the multiplexer flip-flop D-input to zero, disabling the flip-flop and the channel. 

Logarithmic amplifiers make it possible to digitize signals with a wider dynamic range than the digitizer.  The smallest current pulse of interest in an NCD is 200 nA at a gas gain of 100, arising from a proton at the wall oriented toward the anode wire.  The largest current pulse comes from an 8.8-MeV alpha (from $^{212}$Po in the $^{232}$Th decay chain) oriented parallel to the anode wire.  A perfectly parallel track produces a peak current of 29 $\mu$A, limited by lateral straggling of the primary ionization.  Thus the dynamic range required is approximately a factor of 145.  The choice of an 8-bit digitizer and logarithmic amplifier combination keeps digitization noise below the noise floor of the input amplifiers, while still providing GHz digitization speeds.  Higher resolution digitizers available at the time when the design was frozen were either much more costly or capable only of 10-ns samples, comparable to the position resolution desired from the NCDs by pulse reflection.  In a system with a high signal-to-noise ratio, time differences between pulses can be measured much more accurately than the inverse bandwidth, and hence it is important to keep the Nyquist frequency of sampling well above frequencies corresponding to the desired time-difference resolution. In the present system, the pulse-pair time resolution is approximately 5 ns for events with good signal-to-noise ratios, such as alpha particles or neutron tracks parallel to the anode wire.

The offset for the logarithmic amplifiers functions as a gain control and is set via the multiplexer control board.  The offset is also necessary to prevent rectification of the quiescent input noise, as the logarithmic amplifier is indifferent to the sign of the input voltage.  A total of 12 channels shared a single logarithmic amplifier output, so simultaneous events on different channels were read out as a sum of the signals on all channels above their discriminator levels.  The event rate was low enough during normal running (0.3 Hz over 40 channels) that simultaneous events were rare.  The information on which of the 12 channels on a particular multiplexer box exceeded their threshold levels was sent to the multiplexer controller for each digitized event so that the event could be correlated with the appropriate NCD string or strings.  Once the multiplexer controller had triggered the digitization, the data acquisition system received the channel hit pattern for each multiplexer box.  The multiplexers were then rearmed and could receive the next event.  The read-out dead time associated with this was approximately 1 ms.

\subsection{Digitizing oscilloscopes}

Upon receiving notification of a multiplexer channel exceeding its discriminator, the multiplexer control board triggered one of two Tektronix TDS754A oscilloscopes to digitize the output signal from the logarithmic amplifier.  Each of the four channels of the oscilloscope were connected to one of the four multiplexer boxes.  The outputs from the four multiplexer boxes were digitized by the oscilloscope's four channels, but only the signal from the appropriate box (or boxes) was read out by the data acquisition system, which was notified by the multiplexer controller which box and oscilloscope contained the event.  The multiplexer controller triggered the oscilloscope that was not busy (toggling between the scopes when neither was busy), allowing for a maximum digitization rate of 1.8 Hz.  This was adequate for the low data rate during neutrino data-taking but was not sufficient to digitize all events from calibrations or a possible supernova signal.  

The signal was digitized at 1 GHz to provide good vertical position resolution and for pulse-shape discrimination.  For each event, 15 $\mu$s were recorded, including 1500 ns prior to the oscilloscope triggering and about 320 ns of the event prior to the discriminator threshold-crossing, due to the multiplexer delay line.  The remaining 13.2 $\mu$s recorded the ionization event.  The longest duration of interest was the maximum current-pulse width produced by a diametral track, and was approximately 3 $\mu$s.  The long digitization time was convenient to record much of the ion tail caused by the motion of positive ions in the proportional counter, and allowed the signal to return close to the quiescent noise level.

\subsection{Shaper-ADCs}

The NCD signal was also sent from the multiplexer to the input of a custom shaper-ADC channel (one of eight channels on a shaper-ADC card).  The signal was integrated and shaped by a four-stage network of operational amplifiers.  After integration, the shaped signal was split three ways.  It was connected to the analog input of a MAX120CWG 12 bit track-and-hold ADC and to the input of a threshold discriminator.  The shaped signal was also differentiated to form a signal that crossed zero at the peak of the shaped signal, which occurred after about 6 $\mu$s plus one half the pulse width.  A comparator on this zero-crossing clocked a D flip-flop.  The D input was connected to the output of the threshold discriminator such that a flip-flop was set at the peak of the shaper output if the shaped output was greater than the threshold voltage that had been set. The output of this track-hold flip-flop was connected to the `Convert Start' input of the ADC, holding the input voltage and starting conversion.  The track-hold signal also locked out all other shaper-ADC channels after a 180-ns wait to catch all coincident channels.  This signal inhibited any other shaper-ADC events until an embedded CPU (eCPU) resident in the VME crate had read out the triggered channels.  The shaper-ADC system is capable of recording the kHz event rates expected from a galactic supernova.

Each channel also has a scaler counter that counted the number of times the signal crossed the DC-level threshold.  This scaler has a very small dead time (dependent on the size and amplitude of the signal) and was not inhibited at any time.  The scaler counter could be used to determine how many events occurred that were not read out (i.e., missed while the shaper-ADCs were inhibited).  As implemented in the electronics, however, the scaler pulses were derived from zero-crossing signals gated by the threshold comparator and not directly from the logical conversion requests to the ADC. There could occasionally be rapid trains of pulses at threshold (from electromagnetic pickup, for example), for which the scaler counts and the ADC conversions differed by a few counts.  The scalers were thus used only to establish the shaper-ADC dead times with high-rate neutron and random-pulser calibration data well above threshold, but not for shaper dead time correction during normal operation.  The measured shaper dead time (which was enforced for all shapers for each event) is $236 \pm 5$ $\mu$s.

\subsection{Triggering}

Signals from the independent data-acquisition hardware for the NCD array and the PMT array were integrated in a global trigger system that combined both data streams with timing information for the events.  This was accomplished using the SNO Master Trigger Card (SNO-MTC) that assigned a global trigger identification (GTID) number and recorded the times of redundant 10- and 50-MHz clocks for the event.  The SNO-MTC was capable of triggering on a number of external inputs and two of these were used by the NCD trigger system.  The NCD trigger system was controlled by an NCD Trigger Card (NCD-TC) containing a local GTID register that was kept in sync with the SNO-MTC GTID register at all times.  Thus, when an NCD event occurred, the SNO-MTC was informed and assigned a GTID, and this GTID was available for local readout in the NCD data stream.  This allowed the NCD and PMT data to be combined based on the GTID of the event.  The NCD-TC was triggered by the shaper-ADC system or the multiplexer system whenever an event exceeded the threshold on that system.  The NCD-TC resided in the VME crate with the shaper-ADC system and was continuously polled by the VME eCPU.  Once the $\mbox{NCD-TC}$ was triggered, the GTID was latched and read out by the eCPU for use in the data stream.  A local 10-MHz clock on the NCD-TC was also latched and available for read-out and served as a backup to the SNO-MTC clocks.

\subsection{Other electronic systems}

Several other systems were included in the NCD electronics.  These include:

\begin{enumerate}
\item Five Spellman MP3P24 high voltage supplies each capable of providing up to 2300 V to up to ten NCD strings.   The voltages and current draw were controlled and monitored by custom-made high voltage control boards interfaced with the data-acquisition computer.
\item A custom random pulser used to evaluate system dead time.  This pulser is capable of producing an adjustable negative square pulse and a TTL pulse at random times with average rates between 0.0001 Hz and 10 kHz.  This pulser was used continually during neutrino data-taking to provide dead time information, and was set to an average rate of 0.01 Hz.
\item A programmable Agilent 33120A Waveform Generator for use with a custom pulser distribution system.  The pulser distribution system could deliver a pulse to any combination of the NCD preamplifiers to test the electronic functionality, gain, linearity and thresholds of the multiplexer and shaper-ADC systems.  Different pulse shapes could be used for these tasks as well as for checking the logarithmic amplifier properties of the multiplexer system.
\item A set of custom-made live time clocks, one each for the shaper, multiplexer, and oscilloscope systems, that were inhibited when the system is unable to take data while an event is being processed.  An additional clock was never inhibited, providing a means to correct for the dead times of each system.  The clocks could be checked with the random pulser. 
\end{enumerate}

\subsection{Data acquisition system}

The NCD electronics were controlled by the Object-oriented Real-time Control and Acquisition (ORCA) system \cite{howe}.  ORCA is a data acquisition application developed by the University of Washington for the Mac OS X operating system.  The goal of ORCA is to provide a general purpose, highly modular, object-oriented acquisition and control system that is easy to use, develop, and maintain.  ORCA was written using the Mac OS X Cocoa development framework and Objective-C.  The SNO NCDs were the first experiment to utilize the ORCA system.  ORCA was run on a Mac G4 dual processor machine that used an SBS Bit3 620 PCI-to-VME controller to interface with the NCD electronics.  The VME controller directly interfaced with the shaper-ADC boards, the VME eCPU, the NCD-TC, and several Acromag IP408 32-channel digital input/output modules.  Four IP408 modules were used to interface with the multiplexer control board, the high voltage control board, the random pulser, and the pulser distribution system.  The two oscilloscopes were read out via an ENET/100 Ethernet-to-GPIB adapter that also controlled the Agilent pulser used with the pulser distribution system.

ORCA is capable of being issued remote commands from another system.  The SNO Hardware Acquisition and Readout Control (SHARC) software \cite{SNOnim} that serves as the data acquisition for the PMT system could thus remotely start and stop ORCA, keeping run numbers and data acquisition synchronized between the NCD and PMT systems.  Additional control features in SHARC, such as alarm notifications and the issuing of high voltage control commands to ORCA, allowed for SNO detector operators to control both the NCD and PMT systems while interfacing with only one software system.

\subsection{Data streams and monitoring}

The NCD data stream paralleled that of the PMT system, with each stream having its own event builder.  The PMT builder \cite{SNOnim} ran on a Sun~Ultra~5 and read raw data written by the PMT-system front-end eCPU into dual port memory.  The multi-threaded NCD builder ran on a Mac G5 and read the raw ORCA data from the NCD data acquisition machine via a TCP connection with ORCA.  Both builders wrote events to ZEBRA \cite{Zebra} Data Acquisition/Analysis Bank (ZDAB) files.  MD5 checksums were calculated as these primary ZDAB files were being written, and the checksums were carried forward from this point throughout the entire data flow to verify data integrity.

The SNO builder, a multi-threaded second-level event builder also running on the Mac G5, merged the PMT and NCD data streams by reading the primary ZDAB files as they were being written.  The PMT files were read via a dedicated TCP socket connection from a ZDAB server running on the Sun system, while the NCD files were read from the local Mac disk.  Events were combined based on GTID and written to a final SNO ZDAB file.  The SNO builder also sent these events to a network dispatcher that distributed them to the various monitoring and online analysis systems.  Buffering at each stage in the data flow was kept to a minimum, and the final dispatched event was usually available within a few seconds of the original event trigger.  Finally, other data flow processes running on the Mac copied the SNO ZDAB files to Quantum DLT\footnote{Quantum Corporation, San Jose, CA, URL: quantum.com} tapes and sent the files to the surface for additional backup and analysis.

A real-time Qt-based\footnote{Qt is a registered trademark of Trolltech ASA, Oslo, Norway, URL: www.trolltech.com.} program was developed to monitor the NCD array during deployment and operation.  This program could graphically display the data from the NCD builder output files or the real-time output stream from the network dispatcher.  This system for monitoring the NCD array complemented the existing monitoring tools available for the PMTs.

\section{Performance of the NCD array}

\subsection{Signals from the NCDs}
\label{signals}

\begin{figure}[b]
\begin{center} 
\includegraphics[width=\columnwidth]{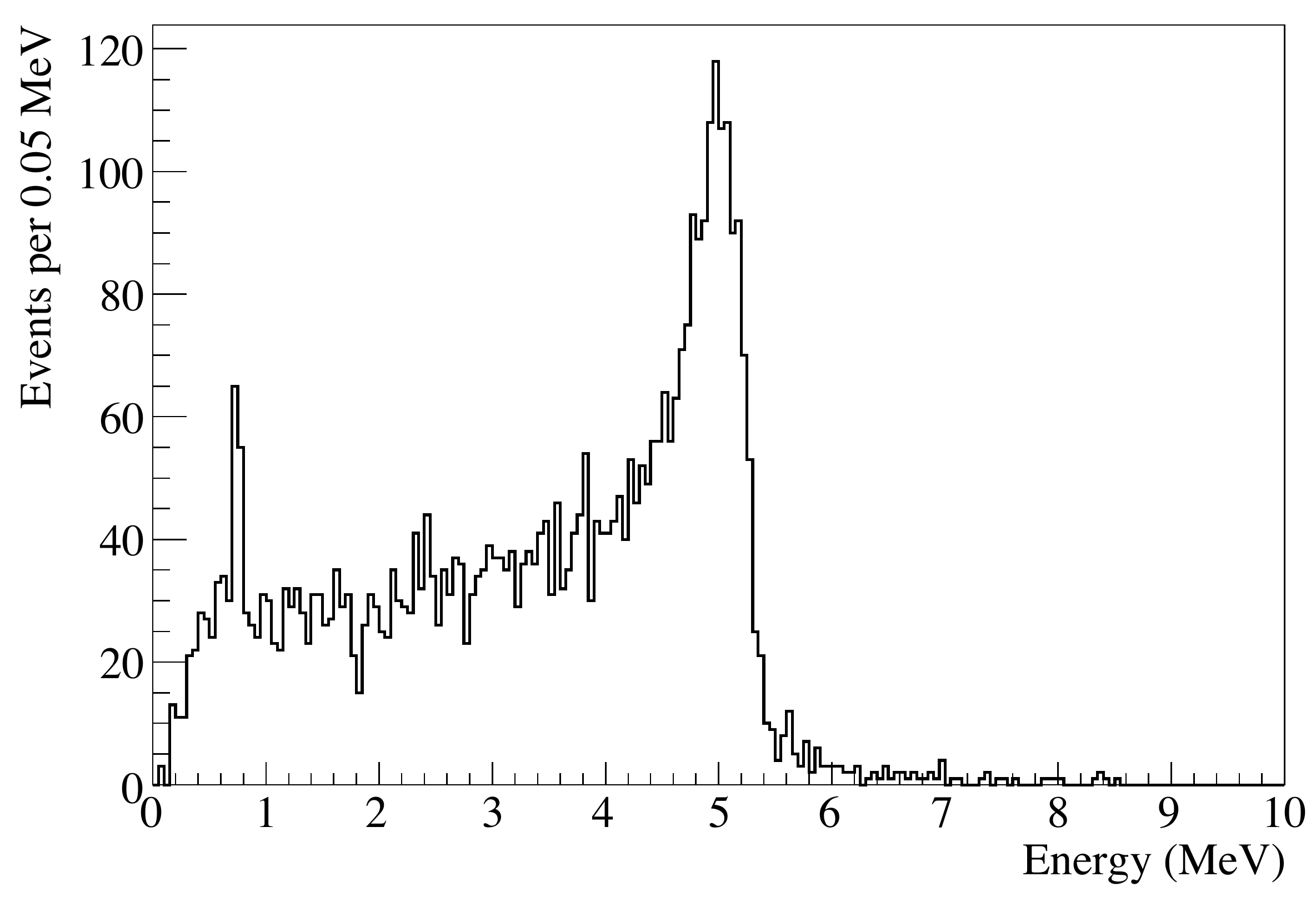}
\caption{Cleaned shaper-ADC energy spectrum from the entire NCD array taken during the initial period of production data-taking.} 
\label{spectAll}
\end{center}
\bigskip
\end{figure}

The primary sources of physics events in the NCDs were neutron captures and alphas from radioactive decays, which produced the energy spectrum shown in Figure \ref{spectAll}.  These shaper-ADC data were taken during the initial period of production data-taking, and have had all the instrumental background cuts described in Section \ref{DC} applied to them.  Slightly over three NC-induced thermalized neutrons per day are predicted to capture on the NCD array.  These, along with additional photodisintegration neutrons, give rise to a neutron-capture energy spectrum below 1 MeV similar to the neutron calibration spectrum shown in Figure \ref{spect}.  Dominating the neutron-capture signal were several hundred alpha events per day arising from natural radioactivity in and on the NCD construction materials.  These were primarily from the decay of $^{210}$Po, a radon daughter that was deposited on the NCD surfaces during construction, and bulk thorium- and uranium-chain activity in the NCD bodies.  The 5.3-MeV surface $^{210}$Po peak is reduced somewhat in energy by space-charge effects, described later in this Section.  The $^{210}$Po peak has a lower-energy tail from alphas emitted at an angle allowing them to strike the NCD wall before depositing their full energy in the active volume.  The bulk activity spectrum below 9 MeV was determined by the energies of the thorium- and uranium-chain alphas, their depth in the NCD bodies, and wall effects.  Energy degradation from the initial depth of bulk alphas and from the wall effect resulted in approximately 20 alphas per day integrated over the whole array that contaminate the neutron-capture energy region.

The current pulse from the NCDs was digitized because pulse-shape discrimination techniques allow for improved separation of alpha backgrounds from the NC neutron-capture signal, compared to just using energy.  The current arriving at the wire is proportional to the rate at which primary ionization drifts in:
\begin{equation}
i=\frac{dE}{dx}\frac{dx}{dr}q\frac{\gamma}{\omega}\left(\frac{dt_d(r)}{dr}\right)^{-1}
\end{equation}
where $\frac{dE}{dx}$ is the specific energy loss along the track, $\frac{dx}{dr}$ describes the geometrical orientation of the track with respect to the wire, $q$ is the charge on the electron, $\gamma$
is the gas gain, $\omega$ is the energy required to create one ion pair, and $\frac{dt_d(r)}{dr}$ is the reciprocal of the effective (radius-dependent) electron drift velocity.  

The primary ionization occurs within a few nanoseconds.  The primary electrons then drift towards the anode wire under the influence of the electric field.  This drift takes about 3 $\mu$s for an electron originating at the cathode wall.  The drift velocity is a property of the gas mixture and pressure, operating voltage, and anode and cathode dimensions, so it could not change from event to event.  Once the primary ionization has drifted to within about 100 $\mu$m of the anode, the electrons are accelerated enough by the strong electric field there to produce an avalanche of secondary ionization proportional to the number of primary electrons.  This avalanche is what provides the gas multiplication factor of the proportional counter, which is about 200 for the NCD gas mixture when operated at 1950 V.  The electron current produced in the avalanche contributes about 2\% of the detected signal, and the rest is due to the secondary positive ions as they slowly drift to the outer cathode wall.  The drift time of the positive ions can be as long as a millisecond, but half of the charge is collected in the first 2 $\mu$s.

Because different particle types have different specific energy loss, the pulse shape conveys information about what caused the event, whether it be proton-triton tracks from a neutron capture, a background alpha, or a spurious pulse.  This information is more difficult to extract when the initial ionization track is nearly parallel to the anode wire, because the total pulse length is very short.  Digitization of the current pulses from the NCDs allows the ionization profile to be deduced, providing event identification for some track orientations.

An important effect that can alter the energy and pulse shapes of proportional counter events is the modification of the electric field in the avalanche region by the presence of positive ions from earlier in the avalanche (or from previous events, although the event rate in the NCDs was small enough that this was negligible).  This effect, known as space charge, can lead to significant reduction in apparent event energy for certain track orientations.  Because the impact of space-charge effects is dependent on the geometry of each event, any space-charge corrections must be geometry-dependent.

\begin{figure}[b]
\begin{center} 
\includegraphics[width=\columnwidth]{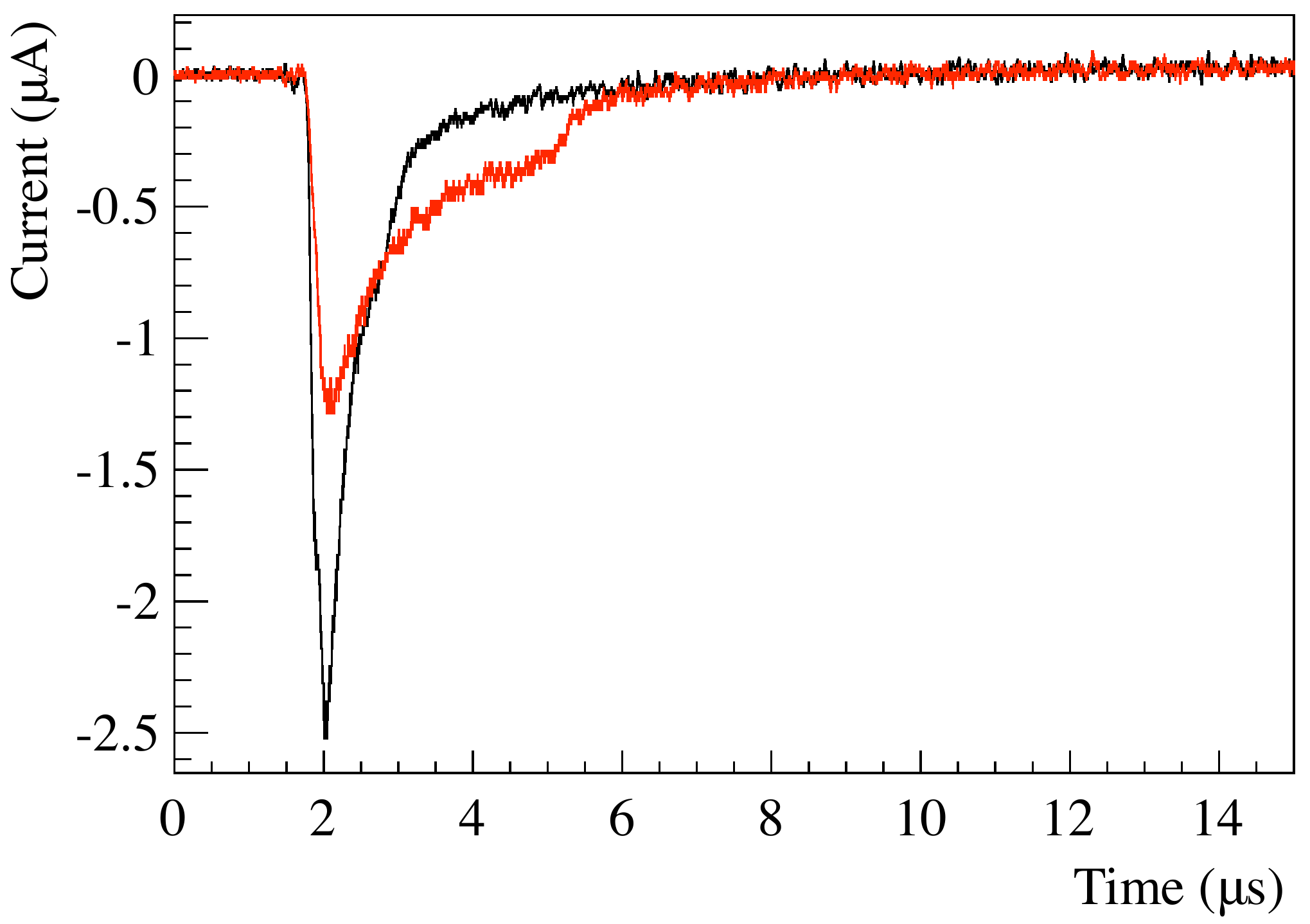}
\caption{Digitized current pulses of two alpha events, each depositing about 4.6 MeV of energy in the active volume of the NCD.  The wider pulse is an alpha event with a track nearly perpendicular to the anode wire, and the narrower pulse had the track nearly parallel to the anode wire.} 
\label{exampleAlphas}
\end{center}
\end{figure}

Two examples of NCD alpha events are shown in Figure \ref{exampleAlphas}.  The wider pulse was produced by an alpha particle that entered the active volume of the proportional counter with an energy of about 4.6 MeV, propagating nearly perpendicularly toward the anode wire.  Near the end of the track, the ionization density from the alpha particle becomes highest (the Bragg peak) and this part of the track arrived first at the wire.  This resulted in a sharp initial maximum in the current followed by a period of decreasing current terminating after about 2 $\mu$s when the most distant primary ionization reached the avalanche region.  In addition to this structure from the drift of the primary electrons, the slow drift of the secondary positive ions produced in each part of the avalanche added a long exponential tail, accounting for the decaying current signal after 2 $\mu$s.  The narrower pulse in Figure \ref{exampleAlphas} was produced by an alpha particle that entered the active volume of the proportional counter with an energy of about 4.6 MeV, propagating nearly parallel to the anode wire.  The pulse is much narrower because the primary ionization reached the avalanche region at nearly the same time from all parts of the track.  The total charge of these events can be measured by integrating the entire pulse, either in software from the digitized pulse or in hardware with the shaper-ADCs.  

\begin{figure}[htb]
\begin{center} 
\includegraphics[width=\columnwidth]{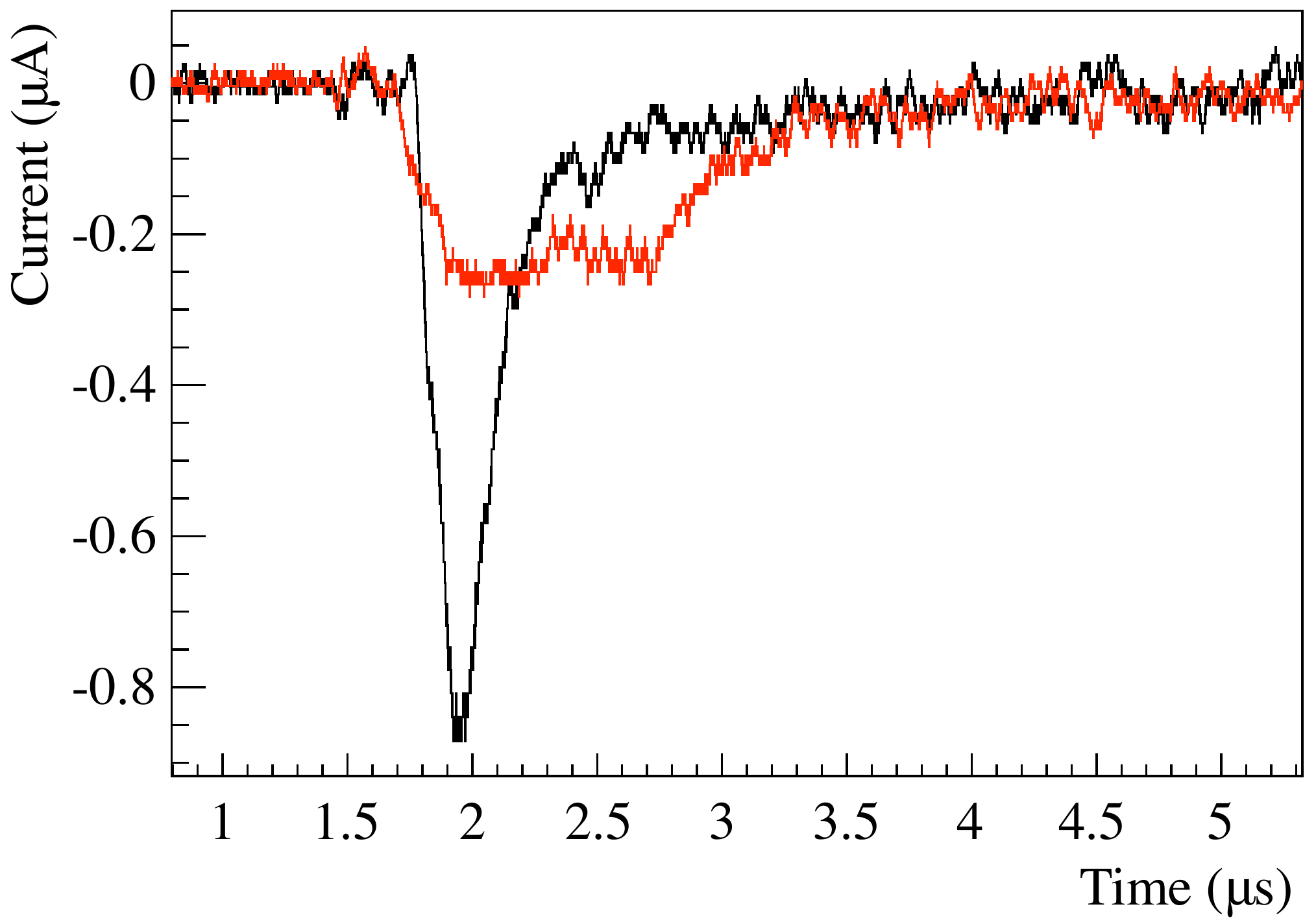}
\caption{Digitized current pulses of two neutron-capture events.  The wider pulse had the proton-triton track nearly perpendicular to the anode wire, and the narrower pulse had the track nearly parallel to the anode wire.} 
\label{exampleNeutrons}
\end{center}
\end{figure}

Figure \ref{exampleNeutrons} shows two neutron-capture events, each with an energy of about 764 keV.  The wider of these pulses had a proton-triton track perpendicular to the anode wire.  In this case, the pulse exhibits a double-peaked structure caused by ionization from the Bragg peak of the proton arriving at the wire first, followed by the rest of the proton ionization, then a discontinuity as the triton ionization began to reach the wire, and then the rest of the triton ionization.  The double-peaked neutron-capture pulse shape is quite different from alpha pulse shapes, and can be distinguished in cases where the proton-triton track is somewhat perpendicular to the wire.  The narrower of these neutron-capture events has a shape very similar to the parallel alpha event shown in Figure \ref{exampleAlphas}.  In general, for these narrow pulse shapes in the neutron-capture energy range, it can be difficult to determine if the track was caused by a neutron or an alpha.  

\subsection{Spurious pulses}
\label{DC}

\begin{figure}[htbp]
\begin{center} 
\subfigure{
\label{nonA}
\includegraphics[width=0.8\columnwidth]{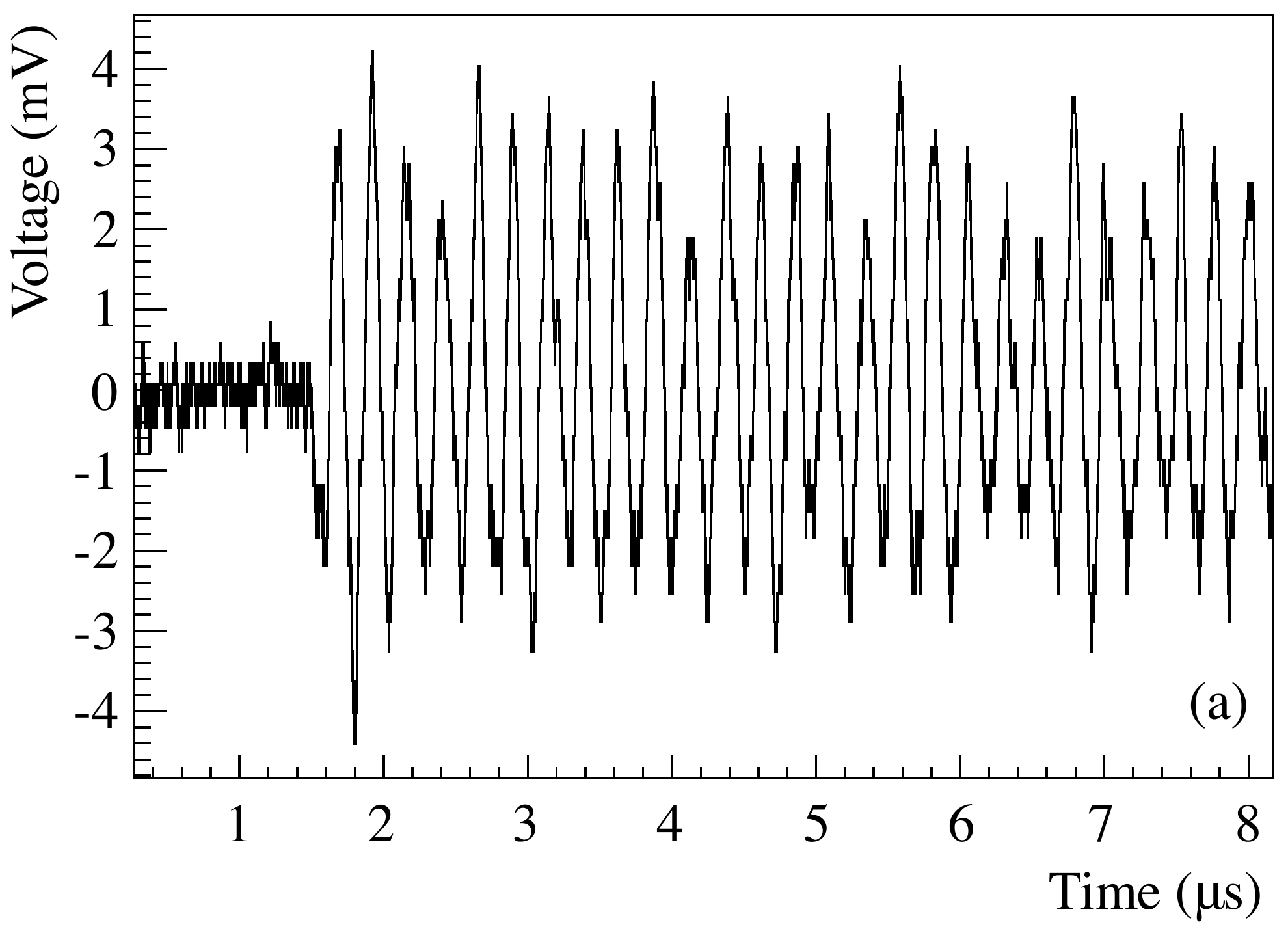}
}
\subfigure{
\label{nonB}
\includegraphics[width=0.8\columnwidth]{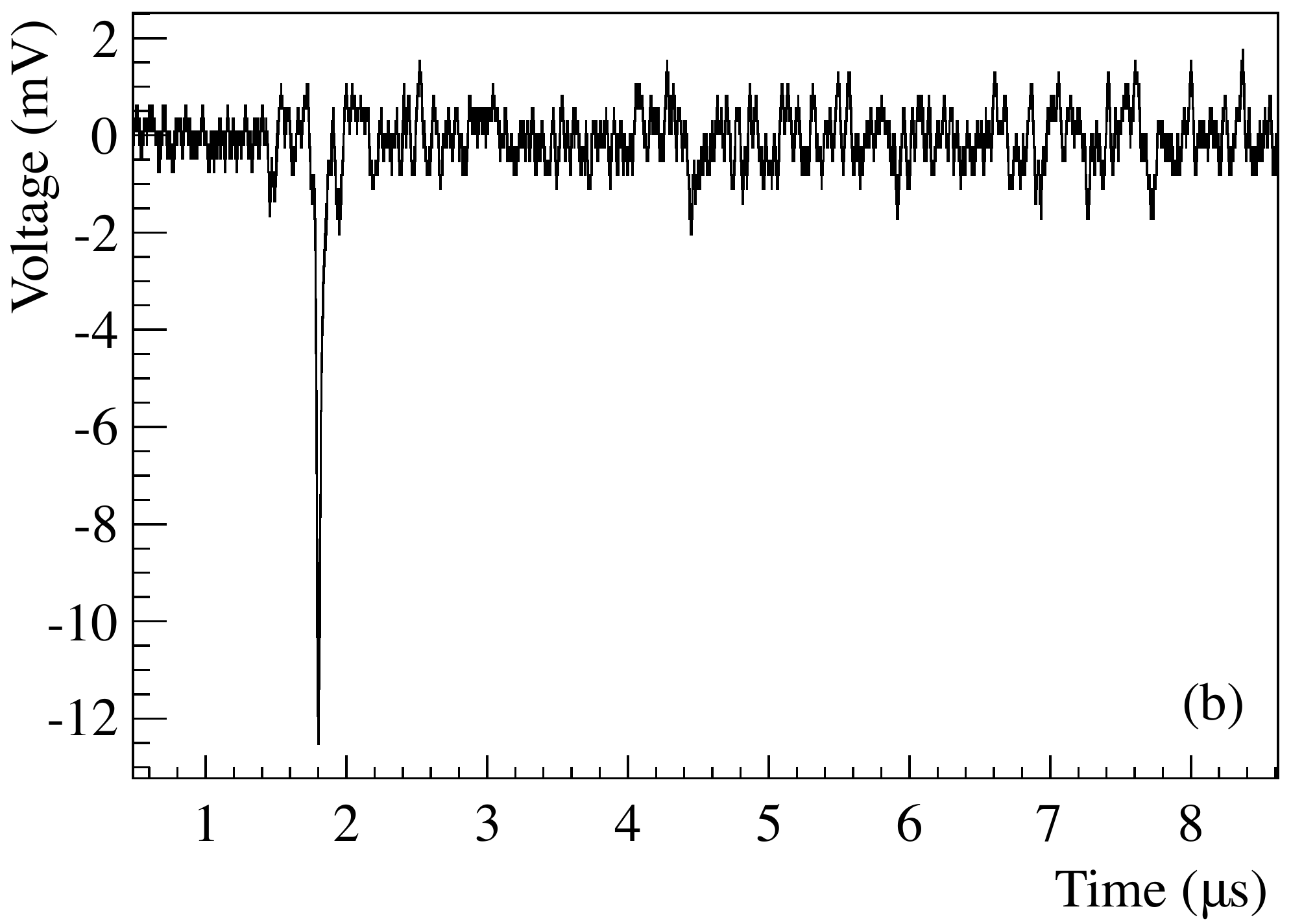}
}
\subfigure{
\label{nonC}
\hspace{-0.15in}
\includegraphics[width=0.83\columnwidth]{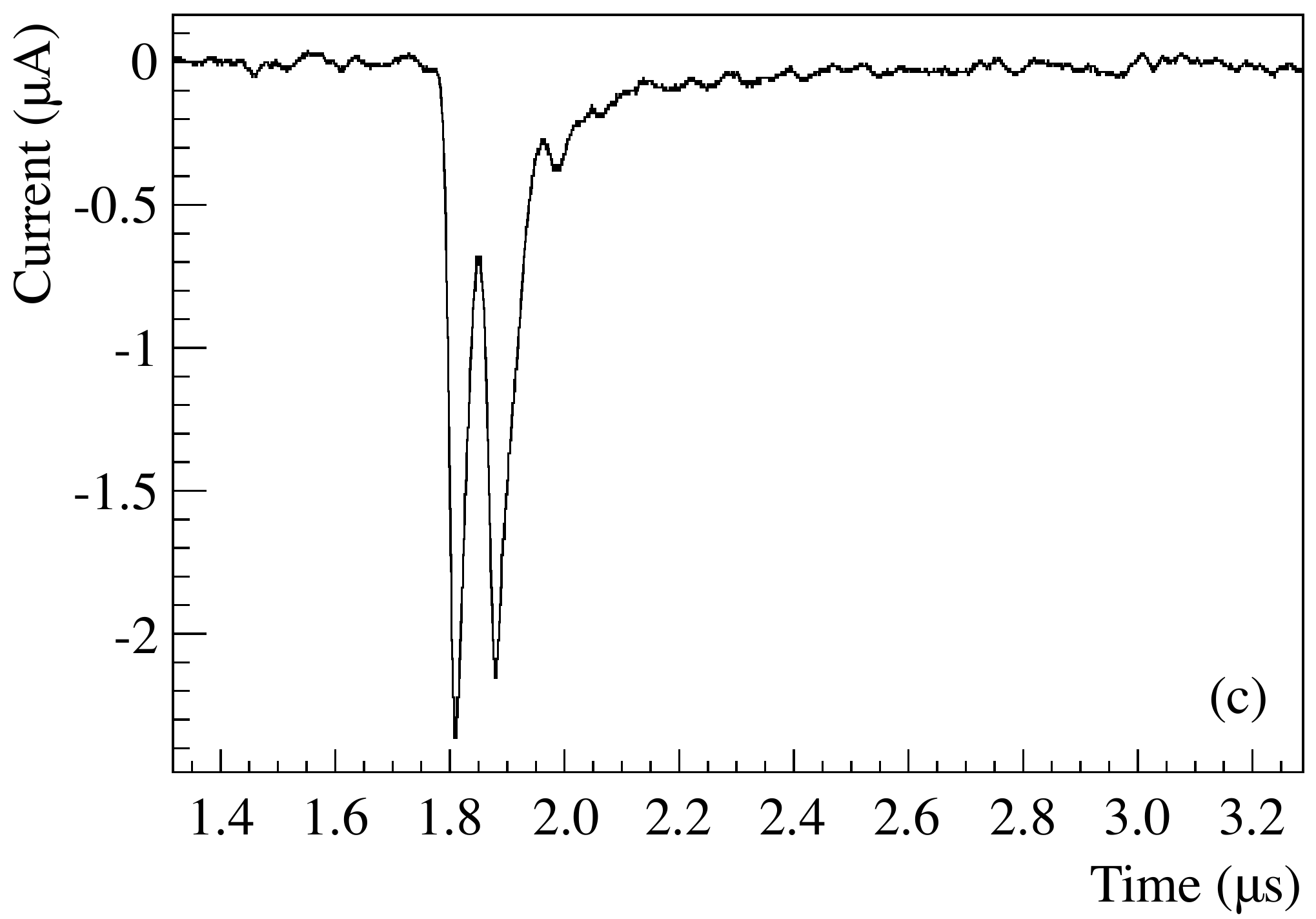}
}
\subfigure{
\label{nonD}
\hspace{-0.2in}
\includegraphics[width=0.84\columnwidth]{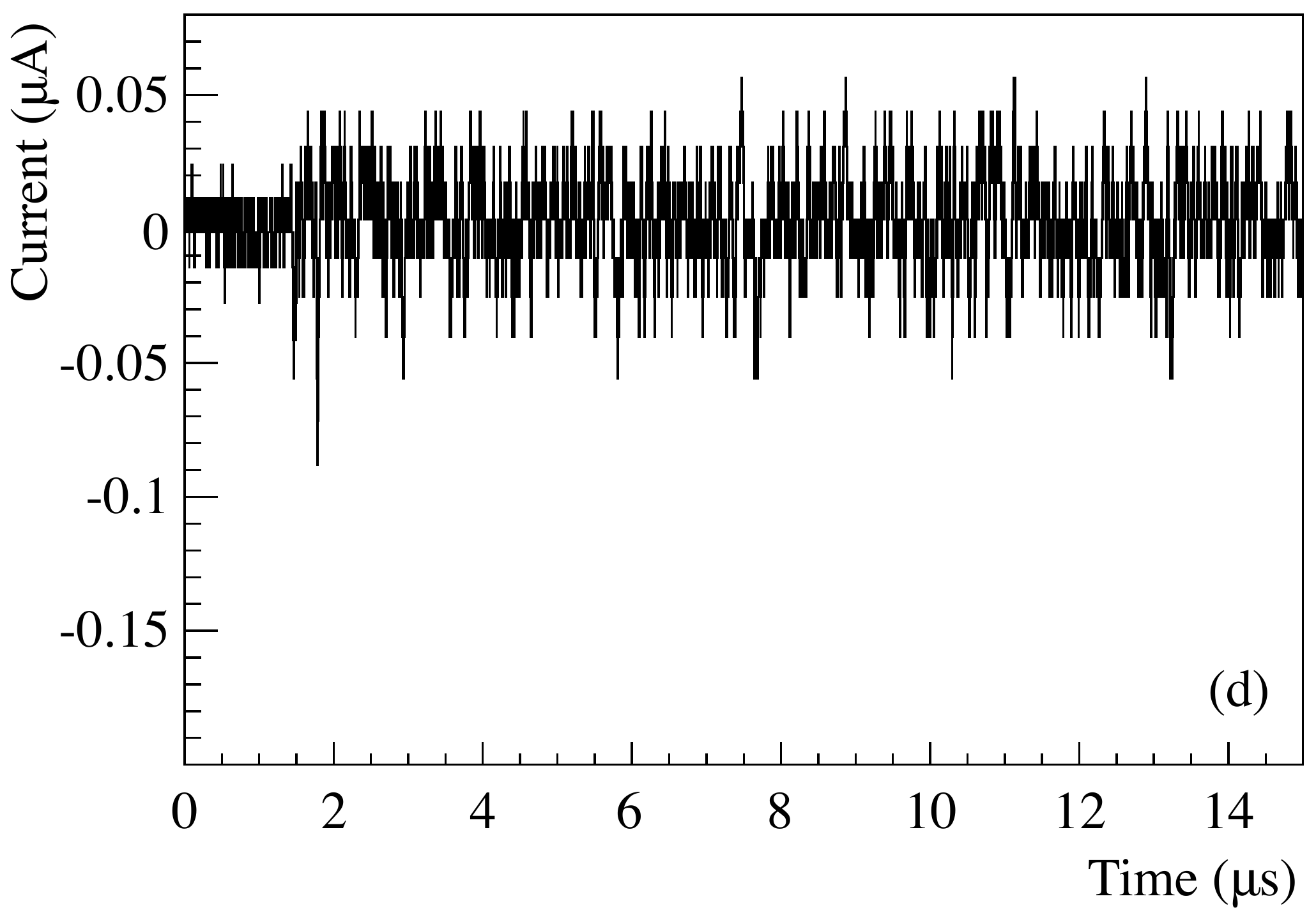}
}
\caption{Examples of non-physics events: oscillatory noise (a), microdischarge (b), microdischarge in the delay line (c), and thermal noise (d).  The events shown in (a) and (b) most likely originated in the preamplifier, so they are shown in units of mV at the preamplifier output, whereas the events in (c) and (d) most likely originated in the NCD string, so they are in units of $\mu$A at the output of the string.} 
\label{non-phys}
\end{center}
\end{figure}

In addition to the physics events in the NCD array, there were a variety of non-physics events that appeared in either the shaper-ADC or the digitized data path, or in both.  The first step in analysis of physics data from the NCDs was the removal of these non-physics events.  Figure \ref{non-phys} shows several examples of these events, including oscillatory noise, microdischarge \cite{microDpaper} in two locations, and a thermal noise event.  The total rate of these non-physics events in the digitized NCD data, which is dominated by thermal noise triggers, was approximately 0.3 Hz.  This is about $10^4$ times higher than the NC neutron detection rate.  However, the rate varied significantly between strings, with many strings having considerably fewer non-physics events.  These pulse shapes can be distinguished from physics events, as described below.

\begin{figure}[htb]
\begin{center} 
\includegraphics[width=\columnwidth]{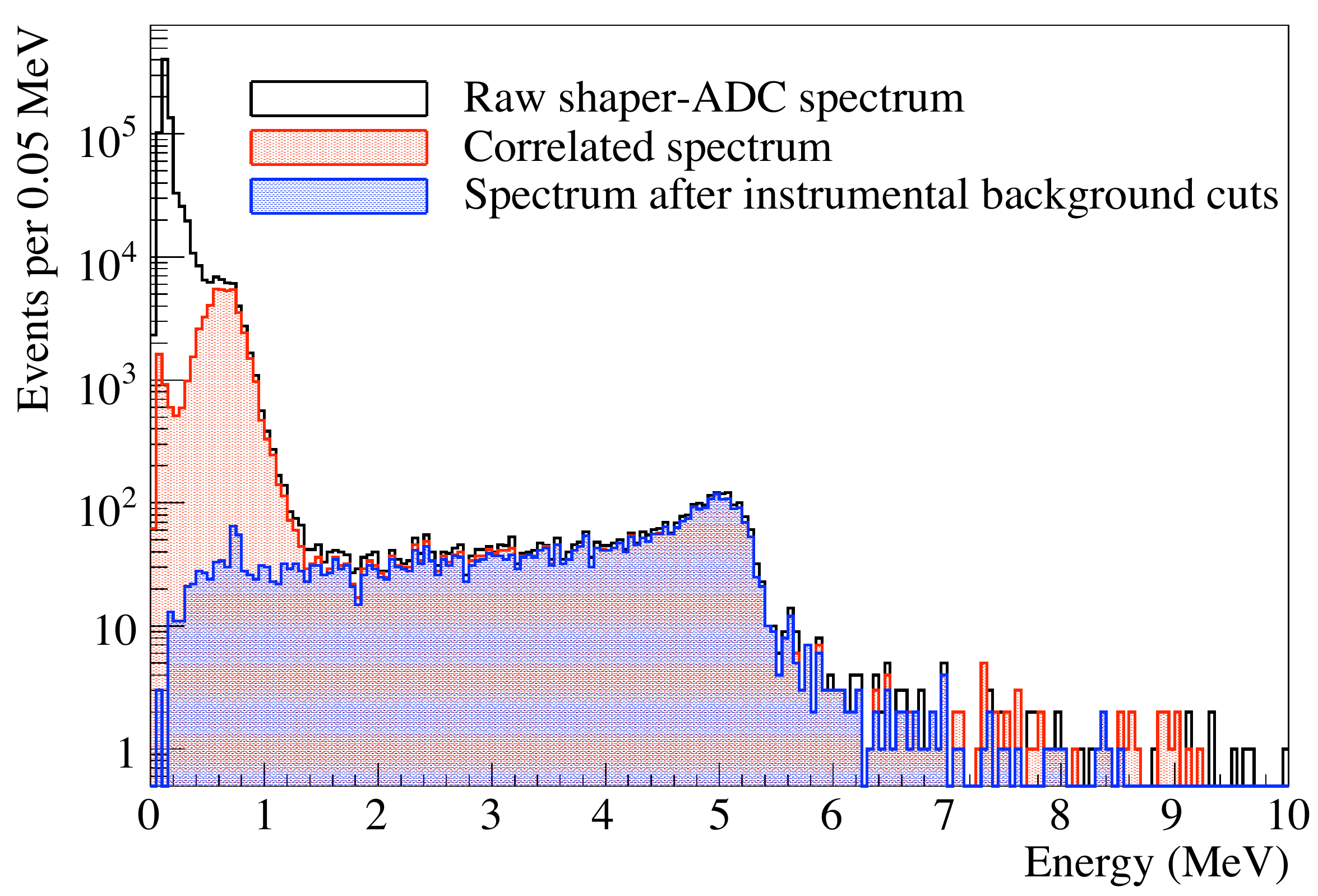}
\caption{NCD array shaper-ADC energy spectra from data taken during the initial period of production data-taking, showing the raw shaper-ADC spectrum, the spectrum of shaper-ADC events with a correlated trigger in the digitized data path, and the shaper-ADC spectrum of events that passed all the instrumental background cuts.  Note that the neutron-capture peak around 764 keV in the cleaned spectrum does not appear very large on this logarithmic scale.} 
\label{dataClean}
\end{center}
\end{figure}

To begin the removal of non-physics events, it was required that both the shaper-ADC and digitized data paths must have triggered on an event.  The shaper-ADC path had a charge trigger and the digitized data path had an amplitude trigger, so some events could be rejected because they did not trigger both data paths.  The impact of this requirement on the shaper-ADC energy spectrum is shown in Figure \ref{dataClean}.  In addition, two independent sets of pulse-shape instrumental background cuts were developed to remove non-physics events from the NCD data.  

One set of instrumental background cuts worked on the logarithmically-amplified digitized waveforms output from the oscilloscopes.  These cuts identified several pulse-shape parameters that differed significantly between physics and non-physics events, and applied cuts on these parameters to remove non-physics events from the data.  For example, extremely narrow pulses such as the one shown in Figure \ref{nonB} were removed by cutting on a pulse width parameter.  Forked events such as the one shown in Figure \ref{nonC} were removed by a cut that parameterizes the relative amplitude of the first and second peak in an event, although care must be taken to avoid cutting physics events with widely-separated initial and reflected pulses.  Other cuts used the frequency of zero-crossings and the RMS noise level in different parts of the event to reject events such as those shown in Figures \ref{nonA} and \ref{nonD}.

The other set of cuts used Fourier transforms of two different regions of the waveform: the pulse region in the first 6 $\mu$s after the trigger, and a noise region 6 $\mu$s long at the end of the pulse that also contains some of the ion tail in physics events.  The power spectrum of the pulse region of each event was divided by the average power spectrum for the noise region for that NCD string, then cuts were applied based on parameters of the power spectrum ratio.  The cuts included a flatness cut that rejected events such as those shown in Figures \ref{nonA} and \ref{nonD}, and a peak frequency cut that removed forked events such as the one shown in Figure \ref{nonC}, as well as other non-physics event types.    

The two sets of cuts were shown to overlap a great deal, but not completely, with 99.46\% percent of cut events removed by both sets of cuts, 0.02\% removed only by the time-domain cuts, and 0.52\% removed only by the frequency-domain cuts.  Less than 1\% of all neutron-capture events were removed by these pulse-shape instrumental background cuts.  Additional instrumental background cuts known as burst cuts removed periods of high-rate data caused by mine activity, HV breakdown in an NCD, noise induced in the NCDs by HV breakdowns in the PMT system, and spallation neutron events following cosmic-ray muons.  The combined effect of all the instrumental background cuts on the shaper-ADC energy spectrum is shown in Figure \ref{dataClean}.  

\subsection{Calibrations and array stability}

A model of the NCD electronics was created for the purpose of estimating the transformation of NCD signals as they propagated through the electronics and were recorded by the shaper-ADC and digitizing oscilloscopes.  The parameters of the electronic transfer function were measured by bench tests of individual electronics components and by a weekly set of in-situ calibrations.  The weekly electronics calibrations (ECAs) tested the linear response of the system, quantified the threshold of each shaper-ADC and digitization channel, and measured the parameters that described the logarithmic amplification and digitization of NCD waveforms.  These measurements were performed by injecting waveforms into each preamplifier's pulser input and observing the measured signal, thereby allowing the linearity, channel thresholds, and pulse-shape transformation to be quantified.  Parameters from each week's ECA were used in analysis of the subsequent week's data to account for small drifts and changes induced by hardware or software modifications.  

\begin{figure}[htb]
\begin{center} 
\includegraphics[width=0.99\columnwidth]{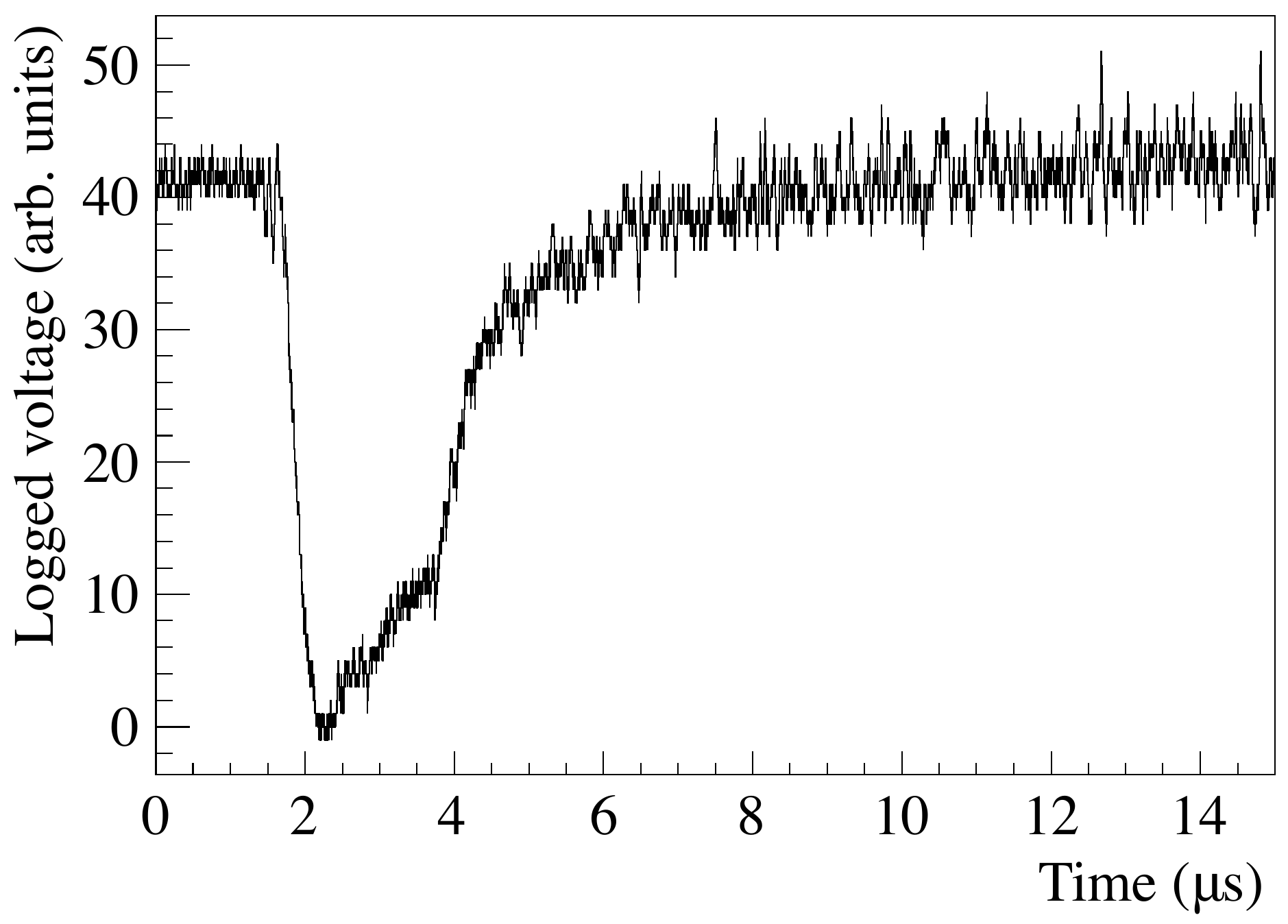}\\
\includegraphics[width=\columnwidth]{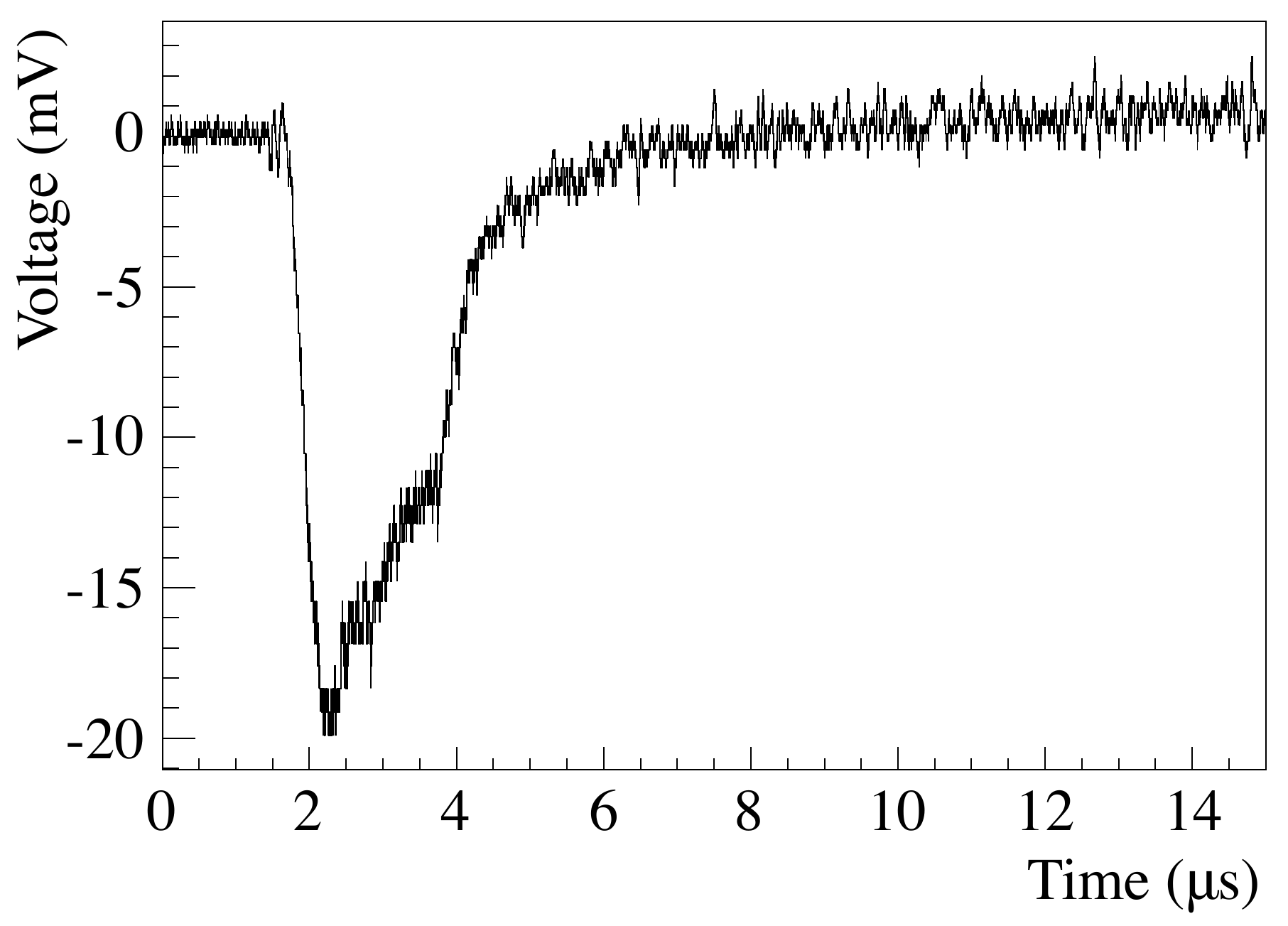}
\caption{An example alpha pulse before and after linearization.  The top panel shows the logarithmically amplified pulse as digitized by the oscilloscopes, with arbitrary logged units on the vertical axis.  The bottom panel shows the same event after linearization, with the vertical axis in mV at the output of the preamplifier.} 
\label{logFig}
\end{center}
\end{figure}

One set of calibrations measured the parameters that describe the effects due to logarithmic amplification and waveform digitization.  Injected pulses were composed of a 78-ns square wave to trigger the data acquisition, followed by an offset, single-cycle sine wave 1.0 $\mu$s in width.  Approximately 30 pulses were collected on each of the 40 channels.  Several parameters describing the amplification and digitization were extracted by fitting the sine-wave portion of the pulses.  These parameters were then used to produce linear waveforms for data analysis, as shown in Figure \ref{logFig}.  Over the two-year span of the NCD phase, the logarithmic amplifier and digitization electronics model parameters changed by less than 3\%.   

The linearity calibration measured the gains and pedestals, and tested the linear response of the system.  Each shaper-ADC channel was pulsed at a rate of 50 Hz for five seconds with 1.1-$\mu$s wide, negative-polarity square waves at five different amplitudes from 36 mV to 155 mV.  Once a month, an extended linearity calibration was performed instead, with 19 different amplitudes from 13 mV to 248 mV.  A linear fit to the shaper-ADC values as a function of the injected pulse charge measured the gain and pedestal of each channel.  The non-linearity was quantified as the difference between the measured charge and the linear fit value and was found to be less than 0.5\% of the measured charge throughout the dynamic range on all channels.  The intercept from the linear equation was then used, along with the peak position in the neutron calibration spectrum, to convert shaper-ADC values to energies.

The threshold calibration extracted the thresholds for each multiplexer and shaper-ADC channel in terms of the input pulse amplitude and total charge.  Offset, single-cycle sine waves 1.0 $\mu$s in width were injected with amplitudes varying from 6 mV to 31 mV, at a rate of 10 Hz for 5 seconds on each channel.  By analyzing which injected pulses triggered the multiplexer and shaper-ADC systems, it was possible to determine the thresholds for each channel.  The thresholds were stable to better than 3\% over the entire NCD phase.

To achieve uniform response across the 36 $^3$He NCD strings, it was necessary to set the thresholds consistently, particularly the multiplexer thresholds.  Low thresholds were desirable to trigger on and digitize all neutron-capture events, however the thresholds could not be set as low as desired due to high noise trigger rates that caused large dead time in the oscilloscopes.  The digitization rate of approximately 0.3 Hz caused the oscilloscopes to be dead about 3\% of the time, which was accounted for in live time calculations.  The multiplexer thresholds miss less than 2\% of all neutron-capture events, and this loss was incorporated into the neutron detection efficiency.

In addition to the electronics calibrations, extensive neutron calibrations were carried out in order to characterize the response of the NCDs to neutrons.  One $^{252}$Cf source (16.55 $\pm$ 0.08 neutrons per second on June 12, 2001) and three $^{241}$AmBe sources ($\sim$ 6 neutrons per second, 23.5 $\pm$ 0.2 neutrons per second, and 66 $\pm$ 2 neutrons per second) were used for these calibrations.  These encapsulated sources were positioned by means of the SNO calibration-source manipulator system \cite{SNOnim}, which allowed 3-dimensional positioning of the source within the SNO AV.  

An isotropic source of neutrons, produced by mixing $^{24}$Na (t$_{1/2}$ = 14.959 hours) into the heavy water on two separate occasions, was also used to calibrate the response of the NCD and PMT systems.  Neutrons were produced by deuterium photodisintegration induced by the 2.754-MeV gammas emitted with the $^{24}$Na beta decay.  The uniformity of this neutron source was ideal, since the NC interaction produces a uniform distribution of neutrons in the heavy water.  However, this calibration was time-consuming because the $^{24}$Na activity must decay away for approximately two weeks to a rate of only a few neutrons per year before solar neutrino data-taking could be resumed.  The efficiency for uniformly-distributed neutrons in the D$_2$O to capture on the NCD array has been preliminarily measured with these calibrations to be $26\% \pm 1\%$.

In addition to measuring the NCD array's neutron-capture efficiency, these neutron calibrations allowed verification that the gain of each NCD string was stable over time by tracking the position of the 764-keV neutron-capture peak.  A leak in the K5 string was found between an active volume and the dead region between sections.  This leak allowed some of the $^3$He-CF$_4$ to exit the active volume and be replaced by air and $^4$He from the dead region, causing a decrease in neutron-capture peak position over many months due to poisoning of the gas mixture.  Because the $^3$He leaked into the dead region rather than into the D$_2$O, this leak did not impact the neutron-transport efficiency of the heavy water.  

To verify that no $^3$He leaked into the heavy water during the NCD phase, a $^3$He detection system was integrated into the existing water recirculation system.  Exhaust gas from the D$_2$O radon-monitor de-gasser system was passed over the outside of heated silica tubes through which helium could readily permeate, then the high-vacuum region inside the silica tubes was measured with a residual gas analyzer to search for a mass-3 signal.  The design sensitivity of this system was the amount of $^3$He that would cause a 1\% change in the neutron-capture cross section of the heavy water (2.2 STP-L of $^3$He diffused throughout the D$_2$O, or about 0.1\% of the total $^3$He in the deployed NCD array).  No mass-3 signal was seen with this system during the NCD phase.  $\mbox{Mass-4}$ signals from a calibration source that diffused $^4$He into the heavy water as part of its normal operation were easily seen, indicating that the system worked.

\section{Radioactivity of the NCD array}

\subsection{Overview}

The NCD radiopurity challenge was to meet two different goals simultaneously: very low thorium and uranium impurity levels to limit the photodisintegration neutron background and also very low rates of alpha backgrounds in the NCDs themselves.  These requirements overlapped somewhat because bulk thorium and uranium contaminants contribute both to intrinsic NCD alpha backgrounds and to photodisintegration neutron backgrounds.  However, $^{210}$Po contamination on the inner surfaces of the NCDs was a source of NCD alpha backgrounds only.  Conversely, thorium and uranium that is not on the inner tens of microns of the NCD wall (contamination on the outer surface, for example) was only a source of photodisintegration neutrons.  

Material selection and careful handling were used to minimize the radioactivity of the NCD array.  Several different techniques were employed to verify that the radiopurity goals were met.  Radioassays of NCD components during construction provided direct measurements of the impurity levels of materials used, but could only be carried out on samples and do not reflect contaminants introduced by later handling of the NCDs.  Analysis of alpha backgrounds in the NCD array provided a good measure of the intrinsic alpha contamination of the NCD neutron signal, but only sampled the inner tens of microns of the nickel walls and had no sensitivity to most other NCD components.  In-situ measurements of Cherenkov light emitted by NCD radioactivity and detected in the PMT array can provide a measure of the overall thorium and uranium content of the NCD array, but are not very sensitive and cannot distinguish well between the two contributions. 

\subsection{Materials controls and radioassays}

\setlength{\tabcolsep}{0.22in}
\begin{table*}[htb]
\bigskip
\caption{Results of radioassays of NCD materials}
\label{radio}
\medskip
\begin{center}
\begin{tabular}{llcccccc}
\hline
\multicolumn{2}{c}{NCD part } & \multicolumn{2}{c}{$^{232}$Th} & \multicolumn{1}{c}{} & \multicolumn{2}{c}{$^{238}$U}\\
\cline{3-4}
\cline{6-7}
 & & Total ($\mu$g) & pg/g & & Total ($\mu$g) & pg/g\\
\hline
\hline
\multicolumn{2}{l}{NCD bodies} & \multicolumn{1}{c}{$0.23 \pm 0.23$} & \multicolumn{1}{c}{$1 \pm 1$} & \multicolumn{1}{c}{} & \multicolumn{1}{c}{$< 1.1$} & \multicolumn{1}{c}{$ < 5$}\\
\hline
\multicolumn{2}{l}{Endcaps} & \multicolumn{1}{c}{$< 1.1$} & \multicolumn{1}{c}{} & \multicolumn{1}{c}{} & \multicolumn{1}{c}{$1.41^{+0.16}_{-0.31}$} & \multicolumn{1}{c}{}\\
             & Nickel sleeves & & $2600 \pm 4400$ & & & $800 \pm 1600$\\
             & Rosin flux & & $2000 \pm 1000$ & & & $3000 \pm 300$\\
             & Flux & & $-700 \pm 500$ & & & $1800 \pm 250$\\
\hline
\multicolumn{2}{l}{Delay line/anchor} & \multicolumn{1}{c}{$< 1.3$} & \multicolumn{1}{c}{} & \multicolumn{1}{c}{} & \multicolumn{1}{c}{$2.02 \pm 0.23$} & \multicolumn{1}{c}{}\\
	    & Delay line boards & & $1100 \pm 400$ & & & $900 \pm 100$\\
             & Anchors & & $130 \pm 230$ & & & $1580 \pm 100$\\
\hline
\multicolumn{2}{l}{Cable bells} & \multicolumn{1}{c}{$< 0.5$} & \multicolumn{1}{c}{} & \multicolumn{1}{c}{} & \multicolumn{1}{c}{$1.84^{+0.07}_{-0.13}$} & \multicolumn{1}{c}{}\\
	    & Heat shrink & & $4100 \pm 800$ & & & $1300 \pm 500$\\
             & Acrylic spacer & & $-74 \pm 141$ & & & $560 \pm 36$\\
             & Coupler & & $-6000 \pm 4000$ & & & $11800 \pm 1400$\\
\hline
\multicolumn{2}{l}{Cables} & \multicolumn{1}{c}{$0.41 \pm 0.47$} & \multicolumn{1}{c}{$90 \pm 102$} & \multicolumn{1}{c}{} & \multicolumn{1}{c}{$2.40 \pm 0.90$} & \multicolumn{1}{c}{$524 \pm 197$}\\
\hline
\end{tabular}
\end{center}
\bigskip
\end{table*}

All materials used to construct the NCD array were selected for low levels of thorium and uranium.  The goal for the NCD array was to limit the thorium and uranium impurities in the NCDs so the photodisintegration background produced by each was less than 1\% of the NC neutron production rate.  This equates to less than 0.5 $\mu$g of $^{232}$Th and less than 3.8 $\mu$g of $^{238}$U in the entire NCD array.  Some NCD components, such as the cables and the anchor assemblies, had less stringent radiopurity requirements because they were concentrated near the AV, so photodisintegration neutrons produced there were more likely to capture on hydrogen in the AV.  Radiopurity specifications for each component of the NCDs were established based on the total mass of that component in the array and its location within the AV.  The nickel NCD bodies had the most stringent radiopurity specification because they dominated the mass of the array and were centrally located in the AV.  

To verify that the levels of thorium and uranium impurities met specifications, material samples and components were radioassayed by radiochemical neutron-activation analysis (RNAA) or by direct counting at the Oroville Dam low-background counting facility operated by Lawrence Berkeley National Laboratory.  Results of these radioassays are shown in Table \ref{radio}.  Most of the parts listed in this table consist of several smaller components.  Each of the smaller components was assayed separately, then the assay results were combined and converted into a total contribution from the entire NCD array (in $\mu$g).  Also shown are individual assay results (in pg/g) for some of the components that contribute the largest total contamination to the NCD parts.  The nickel bodies of the NCDs were not assayed extensively for uranium because the half-life of $^{239}$Np (2.4 days) is inconveniently short to perform RNAA for uranium, and because direct counting does not have the necessary sensitivity for the NCD bodies.  Preliminary RNAA measurements found only upper limits ranging from 1 pg/g to 5 pg/g, indicating that the uranium impurity levels were acceptable.  For all the NCD parts, the total thorium levels were consistent with zero, and in some cases only an upper limit is quoted because many of the components had thorium levels well below the sensitivity of the assay.  The quoted cable impurity levels are only for the portion of the cables below the cable bends, since cables above that point were contained within the neck and produced negligible numbers of photodisintegration neutrons that could capture in the fiducial volume or the NCDs.

\subsection{Radon and $^{\it 210}\!$Po surface contamination}

Another possible source of contamination is radon, which is typically present at the level of 1~pCi/L (37 Bq/m$^3$) in indoor air, and which has been measured at about three times that level in the SNO laboratory \cite{radon}. Radon isotopes and their daughters are all too short-lived to be of much concern for NCD construction, with one exception, $^{210}$Pb from $^{222}$Rn decay. This 22-year activity accumulates on surfaces exposed to radon-laden air, and presents a serious threat to low-background counting experiments.  The low-energy beta decay of $^{210}$Pb leads to $^{210}$Bi (t$_{1/2} = 5$~days), which decays via a 1.2-MeV beta to $^{210}$Po (t$_{1/2} = 138$~days).  The 5.3-MeV alpha from $^{210}$Po produces in a proportional counter a peak with a wall-effect tail that underlies the 764-keV neutron-capture line.

There exists medically-motivated literature on the fate of radon daughters \cite{NN}.  When $^{222}$Rn reaches equilibrium in a space, the concentration is supported against decay and ventilation losses by emanation from construction materials and the ground.  Radon daughters can migrate directly to a surface and be deposited, or they can with much higher probability attach to airborne particulates, which then migrate more slowly before also depositing on surfaces.  To first order, the naive expectation that every radon decay eventually leads to a surface-fixed daughter is borne out.  Based on the `Jacobi Model' \cite{NN}, $^{210}$Po surface activity reaches 1~alpha/m$^2 \cdot$day in under an hour in a typical indoor environment.  

Alpha activity from $^{210}$Po on the inner surface of the NCDs can serve as a continuous calibration, and therefore may not be entirely unwelcome.  For the NCDs, the fraction of surface alpha events in the tail underlying the neutron-capture signal region is at least an order of magnitude lower than for a bulk contaminant, so the relative impact on the neutron signal extraction is much smaller.  In addition, the $^{222}$Rn daughters that attach to surfaces are below $^{214}$Bi in the $^{238}$U chain, so they do not introduce photodisintegration neutron backgrounds.  Despite these facts, it was undesirable to introduce large amounts of surface $^{210}$Po on the NCDs.

As discussed in Section \ref{tubes}, the nickel tubes were inadvertently exposed to high-radon air while stored underground during their construction, necessitating the addition of an electropolishing step to the construction procedures.  After the electropolish, care was taken to minimize additional radon exposure.  The radon level in the University of Washington cleanroom where the detector fabrication took place was found to be below 0.25 pCi/L.  Although this is a low level, several precautions were also taken, including heat-sealing tubes and endcaps in nylon cleanroom bags, minimizing exposure to room air, and excluding air inside the tubes in favor of boil-off nitrogen or vacuum whenever possible.  

Preliminary studies of data from the entire NCD array show $^{210}$Po alpha rates that vary from 0.3~alphas/m$^2 \cdot$day to above 9~alphas/m$^2 \cdot$day on different strings, with an average over the entire NCD array of about 2 alphas/m$^2 \cdot$day \cite{Stonehill}.  However, no correlation is evident between the amount of time (if any) that an NCD tube was stored underground in the high-radon environment during construction and the final $^{210}$Po activity of the NCD produced from that tube.  This suggests that the electropolishing procedure was successful and that small amounts of surface contamination were introduced subsequent to the electropolishing, despite the precautions taken to minimize this exposure.

\begin{figure}[b]
\begin{center}
\includegraphics*[width=\columnwidth]{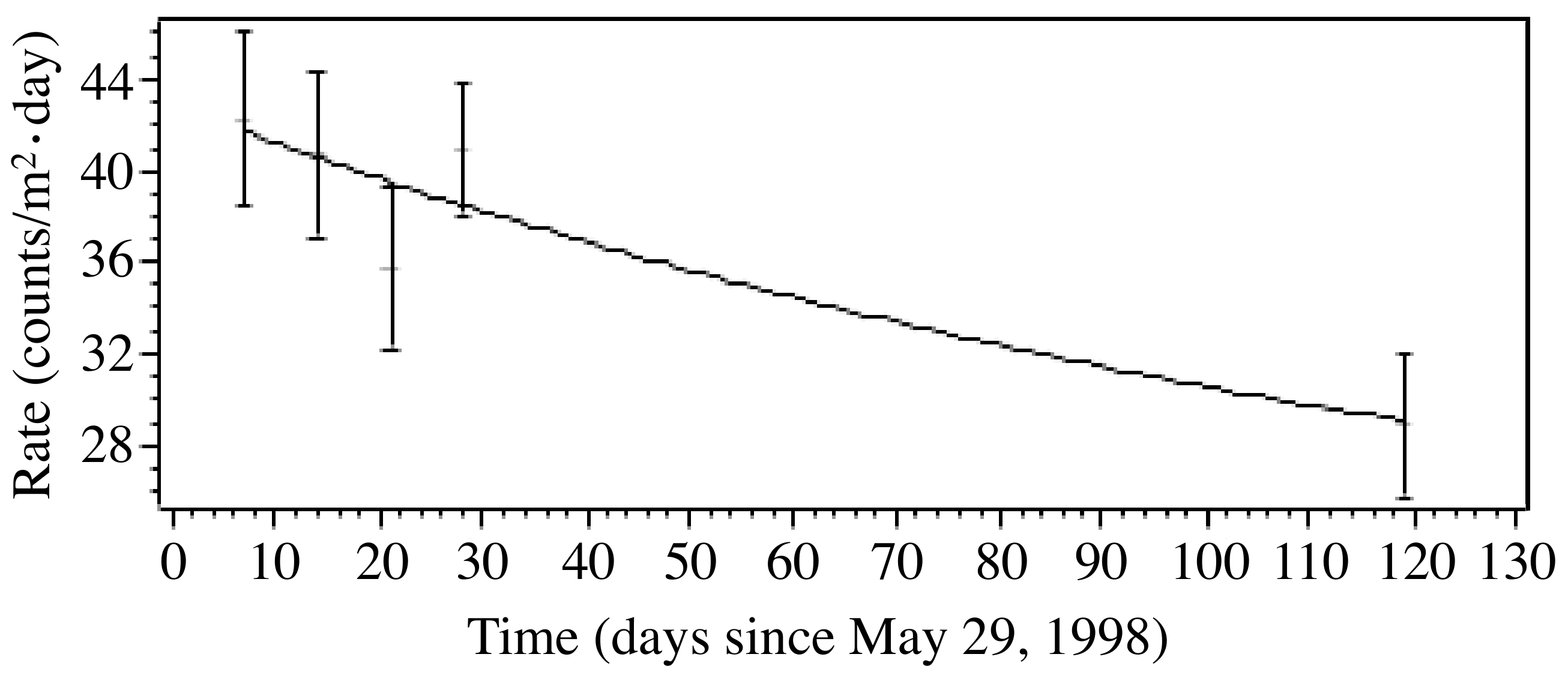}
\caption{\small Count rate recorded in the 4.5 -- 6 MeV energy region from one NCD located in the underground SNO control room between May 29, 1998 and October 5, 1998.  The errors on the data points are statistical only.  The curve is a fit corresponding to a 138-day half-life, appropriate for $^{210}$Po.}
\normalsize
\label{PoDecay}
\end{center}
\end{figure}

Initial measurements of a completed NCD showed $^{210}$Po activity of about 40 alphas/m$^2 \cdot$day between 4.5 MeV and 6 MeV, which was tracked for approximately one $^{210}$Po half-life.  The activity was seen to decay with the 138-day half-life \cite{MCBrowne}, as shown in Figure \ref{PoDecay}.  However, subsequent studies about six years later showed residual $^{210}$Po activity of approximately two alphas/m$^2 \cdot$day of all energies averaged over the entire NCD array, significantly higher than would be expected if the 138-day decay had continued.  Tracking of this activity from June 5, 2004 to January 3, 2005 indicated no measurable decrease in the activity, providing evidence that the remaining $^{210}$Po activity is now supported by 22-year $^{210}$Pb.

\subsection{NCD background measurements}

The primary background to the 764-keV NCD neutron-capture signal arose from alpha decays that did not deposit their full energy in the active volume of the NCD.  In addition to the use of pulse-shape discrimination to distinguish neutron-capture signals from alpha backgrounds, several different techniques were used to measure and characterize the alpha backgrounds in the NCD array.  Studies of the energy spectrum of the alpha events above the region where the neutron signal falls provided constraints on the rates of alphas from different sources.  In addition, a time-coincidence analysis took advantage of short-lived alpha emitters in the thorium and uranium decay chains to provide precise measurements of the levels of these impurities, particularly $^{232}$Th.  

\begin{figure}[b]
\begin{center}
\includegraphics*[width=\columnwidth]{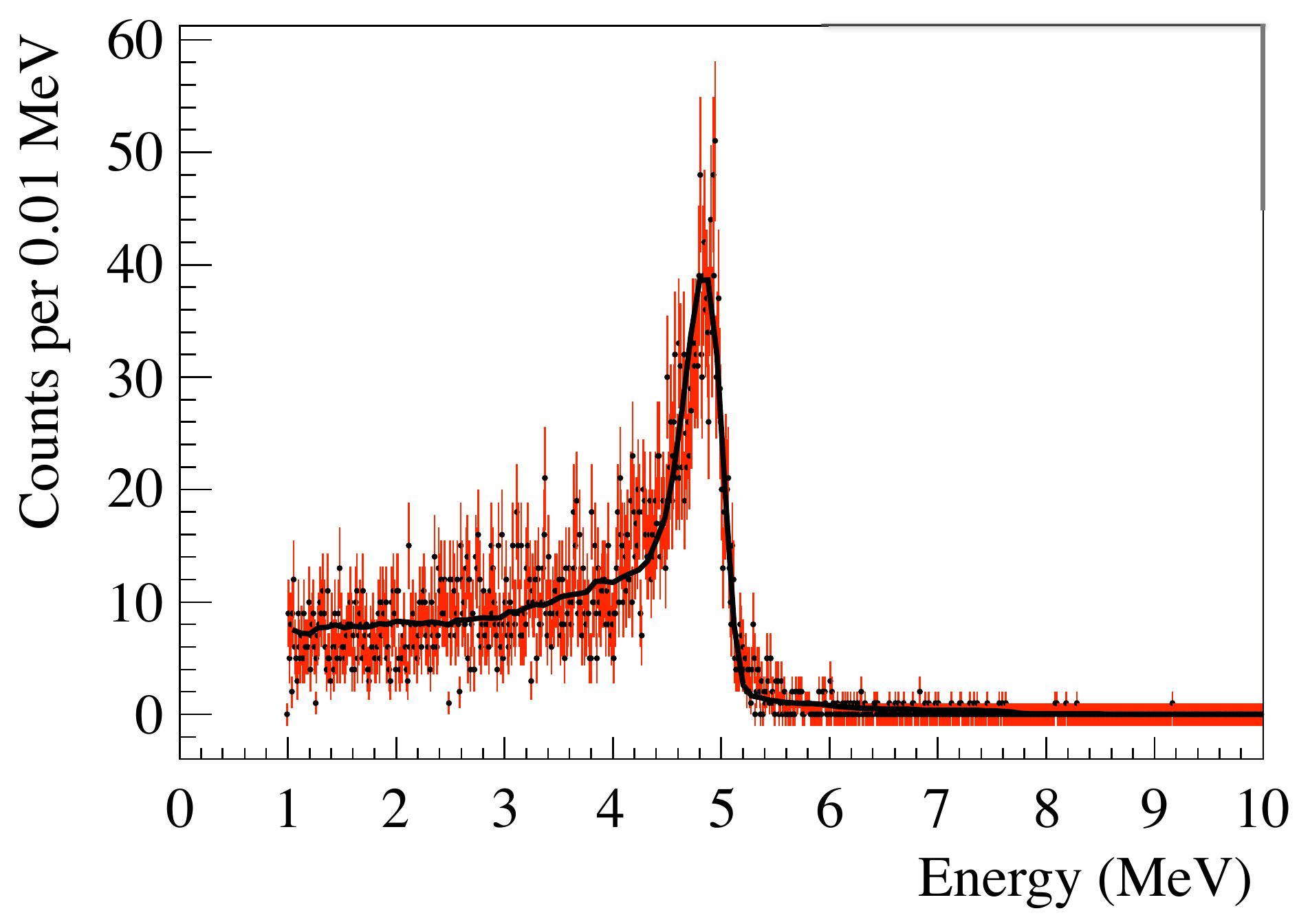}
\caption{\small Combined alpha spectrum above 1 MeV for 36 of the NCD strings from data taken during the initial period of production data-taking.  The peak around 5 MeV is from 5.3-MeV $^{210}$Po alphas, reduced in apparent energy by space-charge effects.  The curve is an unbinned maximum-likelihood fit using alpha energy PDFs based on simulations.}
\normalsize
\label{alphaspec}
\end{center}
\end{figure}

Figure \ref{alphaspec} shows an energy spectrum of events above 1 MeV from data taken during the initial period of production data-taking.  The dominant feature is the 5.3-MeV $^{210}$Po peak, which is reduced somewhat in apparent energy by space-charge effects.  A binned maximum-likelihood fit to the energy spectrum using probability density functions from simulations of surface $^{210}$Po and bulk thorium- and uranium-chain activity is also shown.  The simulations assumed that equilibrium in the thorium and uranium chains had been broken by preferential removal of radium during the CVD process.  The results of this preliminary analysis indicate that the $^{210}$Po activity is $1.9 \pm 0.1$ alphas/m$^2 \cdot$day in the energy region above 1 MeV, and the combined activity from the thorium and uranium chains is $1.6 \pm 0.1$ alphas/m$^2 \cdot$day in the energy region above 1 MeV \cite{Stonehill}.  The energy distribution of the alpha events shows that the amount of surface thorium and uranium contamination on the insides of the NCD tubes is negligibly small.  

By selecting pairs or triplets of alpha events with short time intervals between them, it is possible to identify these events as coming from specific decays in the thorium and uranium chains, with low rates of accidental coincidences.  In the uranium chain, the half-life of $^{218}$Po is 3.05 minutes.  In the thorium chain, the half-life of $^{220}$Rn is 55.6 seconds and the half-life of $^{216}$Po is 0.15 second, thus it is possible to observe the triple coincidence between the alphas from $^{224}$Ra, $^{220}$Rn, and $^{216}$Po.  The result of this preliminary analysis on data taken during NCD commissioning and the initial period of production data-taking was $0.5 \pm 0.1$ $^{232}$Th decays/m$^2 \cdot$day and $0.5 \pm 0.1$ $^{238}$U decays/m$^2 \cdot$day \cite{Stonehill}.  This result can be compared to the $0.3 \pm 0.1$ $^{232}$Th decays/m$^2 \cdot$day and $0.8 \pm 0.2$ $^{238}$U decays/m$^2 \cdot$day obtained from the energy spectrum fits described above.  

The thorium activity measured by these two analysis techniques is equivalent to a $^{232}$Th impurity level of $6 \pm 1$ pg/g \cite{Stonehill}, which is about six times higher than the $1 \pm 1$ pg/g measured in the CVD nickel radioassay.  The uranium activity measured with these alpha analyses is equivalent to a $^{238}$U  impurity level of $3 \pm 1$ pg/g, consistent with the upper limit of 5 pg/g obtained from the preliminary uranium radioassays performed on the NCD nickel.

By using simulation to project these measured alpha rates below 1 MeV, the total alpha rate in the neutron-capture energy range from 155 keV to 800 keV was calculated to be $16 \pm 1$ alphas per day in the 36 $^3$He NCD strings \cite{Stonehill}, about five times the signal expected from neutrons produced by neutral-current interactions.  Although this signal-to-background ratio is not favorable, the distinctive shape of the neutron-capture energy spectrum can be used to extract the neutron-capture signal from the alpha background to an accuracy of approximately 5\% using energy information alone.  Taking advantage of the additional information available in the digitized pulses is expected to improve separation of the neutron-capture signal.  

\subsection{PMT measurements}

Photodisintegration induced by $^{208}$Tl and $^{214}$Bi gammas was the primary source of neutron background in SNO.  In the previous phases of SNO, trace amounts of $^{232}$Th and $^{238}$U daughters in the D$_2$O were the dominant source of this neutron background and this was quantified by ex-situ and in-situ techniques \cite{longd2o, salt2}.  Ex-situ measurements involved regular assays of the D$_2$O using three complementary techniques \cite{HTiO, MnOx, radon}. The in-situ technique measured the Cherenkov light from $^{208}$Tl and $^{214}$Bi decays and used the difference in the isotropy of the light from the events to measure the amount of each isotope present.  Only the in-situ technique could be used to measure the additional photodisintegration background originating from the NCD array.

To separate Cherenkov signals originating on the NCDs from those originating in the D$_2$O, the difference between their radial distributions was exploited.  By placing cylindrical cuts around each NCD string, the radial profile of each background component could be studied and the origin of the Cherenkov light deduced.   Preliminary studies indicate that it is possible to separate activity originating in the D$_2$O from that on an NCD, but it is not possible to differentiate between the two NCD components, $^{208}$Tl and $^{214}$Bi.  The isotropy of the light seen in the PMTs, which was typically used to distinguish between $^{208}$Tl and $^{214}$Bi, could not be used for NCD activity because the betas were attenuated by the nickel body of the NCD.   Almost all $^{208}$Tl decays emit a gamma that is capable of photodisintegration, compared to $^{214}$Bi where the probability is around 2.14\%.    However, since only the gammas escape into the water to be detected, the lack of knowledge about whether they came from $^{238}$U or $^{232}$Th turns out to be less important.  The uncertainty in the neutron photodisintegration rate caused by activity within the NCDs is less than a factor of two, rather than the factor of 50 suggested by the branching ratios.  Assuming that all of the NCD activity is thallium-like provides a worst-case estimate of the photodisintegration background rate that could be a factor of two too large.  This estimate gives a $^{232}$Th impurity level comparable to that measured with the NCD alpha analyses.  The photodisintegration neutron rate from this amount of $^{232}$Th is expected to be approximately 0.25 neutron produced per day. 

In addition, the in-situ analysis is sensitive to `hot spots' of increased localized activity on part of an NCD.  Two such hot spots have been identified, one on K2 and one on K5.  Preliminary studies of the hotter of the two, on K5, indicate that it is expected to produce no more than a few percent of the NC neutron rate, and that most of these photodisintegration neutrons recaptured on the K5 string.  There is evidence that this hot spot was caused by external surface activity introduced during deployment of this NCD string.  Alpha counting and destructive assays of this hotspot are underway in an attempt to identify the composition of this hot spot, with the intent of reducing the uncertainty on the photodisintegration background it produces.

\subsection{Comparison to other detectors}

The preliminary NCD nickel impurity measurements of $6 \pm 1$ pg $^{232}$Th/g and $3 \pm 1$ pg $^{238}$U/g are much lower than the levels in commercially-available $^3$He proportional counters, or even any $^3$He proportional counters designed for previous low-background applications.  Typical low-background $^3$He proportional counters are concerned only with alpha contamination of the neutron-capture signal, and not also with photodisintegration neutron production, as in the case of the SNO experiment.  Thus the region of concern for contamination is only the inner tens of microns of the proportional-counter wall, so applying a low-background coating to the inner wall is an effective background-reduction technique.  

A group at University of California at Irvine prepared copper-coated stainless steel $^3$He proportional counters that reduced the alpha rate from $3 \times 10^3$~alphas/m$^2 \cdot$day with untreated stainless steel to $8 \times 10^2$~alphas/m$^2 \cdot$day with the electroplated copper coating \cite{Pasierb}.  A Soviet group coated the inner surface of stainless steel $^3$He proportional counters with 60 microns of an ``organofluorine compound having an activity 10 times less than the steel'' and observed rates of about $10^2$~alphas/m$^2 \cdot$day \cite{Soviet}.  The NCD rate of 4 alphas/m$^2 \cdot$day across the entire spectrum is orders of magnitude below the rates achieved by these previous detectors.  

\section{Conclusions}

The Sudbury Neutrino Observatory's ability to measure the total flux of all active neutrinos from the Sun above 2.2 MeV via neutral-current interactions was enhanced by the addition of the NCDs.  These 40 strings of $^3$He and $^4$He proportional counters occupied the central region of the SNO acrylic vessel on a 1-meter grid.  The neutron-capture efficiency of the NCD array has been preliminarily measured to be $26\% \pm 1\%$ for uniformly distributed neutrons.  The NCD signal was completely independent of the PMT array, allowing NC neutrons that captured in the NCDs to be identified separately from CC and ES signals in the PMTs.

The main consideration in the design of the NCD array was the reduction of radioactive impurities.  Photodisintegration induced by $^{208}$Tl and $^{214}$Bi gammas was the primary source of neutron backgrounds to a NC flux measurement in SNO, and thus strict controls of all thorium and uranium contamination were required.  Additionally, the primary background to the 764-keV neutron-capture signal was alpha decays that did not deposit their full energy in the active volume of the NCD, and thus any alpha-producing impurities on or near the inner surface of the NCD bodies were also a concern.  A variety of innovative materials and techniques were used to construct and install the NCDs, and extensive testing and optimization was carried out before, during, and after deployment of the NCD array to ensure technical feasibility and data quality.  

Preliminary measurements indicate that the photodisintegration neutron background produced by bulk thorium and uranium impurities in the NCD array was about 2\% of the NC neutron rate.  Neutron backgrounds produced by NCD surface hot spots have been limited to no more than a few percent of the NC rate.  The photodisintegration neutron background produced by the NCD array appears to be slightly higher than originally anticipated, but additional analyses of the hot spots are underway with the goal of reducing the systematic uncertainties associated with this background to within acceptable limits.  

A total observed rate of less than 4 alphas/m$^2 \cdot$day for the entire array of 63.52 m$^2$ has been achieved.  In the neutron-capture energy region of interest below 1 MeV, the total alpha rate was $16 \pm 1$ per day in the $^3$He portion of the array.  This is an improvement of orders of magnitude over previous low-background proportional counters.  Preliminary studies of the efficacy of the pulse-shape analysis techniques under development indicate that the alpha rates in the NCD data can be significantly reduced with a small loss of the neutron-capture signal.  

Data-taking in the NCD phase of SNO began in November 2004 and ended in November 2006.  A blind analysis of the data acquired is underway.  Overall, the NCD-phase goal of creating a robust array of low-background $^3$He proportional counters capable of taking data for several years while submerged in ultrapure heavy water has been met.  

\section*{Acknowledgments}

This work was supported primarily by the United States Department of Energy Office of Science and by Los Alamos Laboratory-Directed Research and Development.  Additional support was provided in Canada by Natural Sciences and Engineering Research Council, Industry Canada, National Research Council, Northern Ontario Heritage Fund, Atomic Energy of Canada, Ltd., Ontario Power Generation, and Canada Foundation for Innovation; in the US by National Energy Research Scientific Computing Center and Alfred P. Sloan Foundation; and in the UK by Particle Physics and Astronomy Research Council.  We thank the SNO technical staff for their strong contributions, and CVRD Inco, Ltd. for hosting the SNO experiment.

\end{document}